\documentclass[11pt]{article}
\usepackage{epsfig,amsmath,amssymb,euscript}
\usepackage[square,comma,sort&compress,numbers]{natbib}
\usepackage{color}
\usepackage{graphicx}
\usepackage{lineno}

\begin{document}
\title{On the Design and use of Ensembles of Multi-model Simulations for Forecasting}


\author{Sarah Higgins$^{1}$, \quad \quad Hailiang Du$^{1,3}$ and \quad \quad  Leonard A. Smith$^{1,2,3}$ \\[2ex]
        $^1$Centre for the Analysis of Time Series,\\
       London School of Economics, London WC2A 2AE. UK\\[1ex]
   $^2$Pembroke College, Oxford, UK \\[1ex]
   $^3$Center for Robust Decision Making on Climate and Energy Policy,\\
   University of Chicago, Chicago, IL, US
   }

\date{\today}

\maketitle

\begin{abstract}
Probability forecasting is common in the geosciences, the finance sector, and elsewhere. It is sometimes the case that one has multiple probability-forecasts for the same target. How is the information in these multiple forecast systems best ``combined"? Assuming stationary, then in the limit of a very large forecast-outcome archive, each model-based probability density function can be weighted to form a ``multi-model forecast" which will, in expectation, provide the most information. In the case that one of the forecast systems yields a probability distribution which reflects the distribution from which the outcome will be drawn, then Bayesian Model Averaging will identify this model as the number of forecast-outcome pairs goes to infinity. In many applications, like those of seasonal forecasting, data are precious: the archive is often limited to fewer than $2^6$ entries. And no perfect model is in hand. In this case, it is shown that forming a single ``multi-model probability forecast" can be expected to prove misleading. These issues are investigated using probability forecasts of a simple mathematical system, which allows most limiting behaviours to be quantified.

\end{abstract}

\section{Introduction}

Forecasters are often faced with an ensemble of simulations which are to be interpreted as making quantitative predictions. Indeed, ensembles of initial conditions have been operational in weather centers in both America~\cite{Kirtman13} and Europe~\cite{Palmer04,Weisheimer} since the early nineties and there is a significant literature on their interpretation~\cite{Raftery05,Hoeting99,Roulston03,Wang04,Wilks06,Wilks07}. There is significantly less work on the design and interpretation of ensembles over model structures, although such ensembles are formed on weather (TIGGE~\cite{Bougeault10}), seasonal (ENSEMBLES~\cite{Weisheimer}) and climate (CMIP5~\cite{Taylor12}) forecast lead-times. This paper focuses on the interpretation of multi-model ensembles in situations where data are precious, that is where the forecast-outcome archive is relatively small. Archives for seasonal forecasts fall into this category, typically limited to between 32 and 64 forecast-outcome pairs.\footnote{{The observational data available for initialization of the forecasts is very different before the satellite era.}} At times the forecaster has only an ``ensemble of convenience" composed by collecting forecasts made by various groups for various purposes; alternatively, multi-model ensembles could be formed in collaboration using an agreed experimental design. This paper was inspired by the ENSEMBLES project~\cite{Weisheimer}, in which seven seasonal models were run in concert, with nine initial condition simulations under each model~\cite{Hewitt04}. Small archive parameters (SAP) forecast systems are contrasted with large archive parameters (LAP) forecast systems using the lessons learned in experimental design based on the results originally reported by Higgins~\cite{Higgins15}. This is illustrated using a relatively simple chaotic dynamical system. Specifically, the challenges posed when evaluation data are precious are illustrated by forecasting a simple one-dimensional system using four imperfect models. A variety of ensemble system designs are considered: the selection of parameters, including model weights, and the relative value of ``more" ensemble members from the ``best" model are discussed. In the large forecast-archive limit, the selection of model weights is shown to be straightforward and the results are robust; when a unique set of weights are not well defined, the results remain robust in terms of predictive performance. It is shown that when the forecast-outcome archive is nontrivial but small, as it is in seasonal forecasting, uncertainty in model weights is large. {The parameters of the individual model probability forecasts vary widely in the SAP case; they do not in the LAP case. This does not guarantee that the forecast skill of SAP is significantly inferior to that of LAP, but it is shown that in this case the SAP forecast systems are significantly (several bits) less skillful. The goal of this paper is to refocus attention on this issue; not claim to have resolved it given only SAP is made. Turning to the question of forming a multi-model forecast system, it is shown that (a) the model weights assigned given SAP are significantly inferior to those under LAP (and, of course, to the using ideal weights).} (b) Estimating the best model in SAP is problematic when the models have similar skill. (c) Multi-model ``out of sample" performance is often degraded due to the assignment of low (zero) weights to useful models. Potential approaches to this challenge, beyond waiting many decades, are discussed.

%
%
%
%

\section{From Ensemble(s) to Predictive Distribution}

The ENSEMBLES project considered seasonal forecasts from seven different models; an initial condition ensemble of 9 members was made for each model and launched four times a year (in the months of February, May, August and November). The maximum lead time was 7 months, except for the November launch which extended to 14 months. Details of the project can be found in~\cite{Alessandri,Doblas10,Weigel,Hewitt04,Weisheimer,Smith14}

The models are not exchangeable in terms of the performance of their probabilistic forecasts. Construction of predictive functions via kernel dressing and blending with climatology (see ~\cite{Silverman86,Brocker08} and Appendix A) for each initial condition ensemble of simulations is discussed in~\cite{Smith14} (under various levels of cross validation). Throughout this paper, IJ Good's logarithmic score is used~\cite{Good52,Roulston02}, this score is sometimes referred to as Ignorance (IGN)~\cite{Roulston02}. IGN is the only proper and local score for continuous variables~\cite{Bernardo79,Raftery05,Brocker06}, it is defined by:
\begin{eqnarray}
 \label{eq:IGN}
    S(p(y),Y)=-\log_{2}(p(Y)),
\end{eqnarray}
where $Y$ is the outcome and $p(Y)$ is the probability of the outcome $Y$. In practice, given $K$ forecast-outcome pairs $\{(p_{i},Y_{i})\mid i=1,\dots,K\}$, the empirical average Ignorance score of a forecast system is then  
\begin{eqnarray}
 \label{eq:eIGN}
    S_{E}(p(y),Y)=\frac{1}{K}\sum_{i=1}^{K}-\log_{2}(p_{i}(Y_{i})),
\end{eqnarray}

%

\section{Simple Chaotic system models pair}

Without any suggestion that probabilistic forecasting of a one-dimensional chaotic map reflects the complexity or the dynamics of seasonal forecasting of the Earth System, this paper draws parallels between challenges to probability forecasting of scalar outcomes using multiple models with different structural model errors when the forecast-outcome archive from the system is small; these challenges occur both in low dimensional systems and in high dimensional systems. Whether or not suggestions inspired by the low-dimensional case below generalise to high dimensional cases (or other low dimensional cases, for that matter), would have to be evaluated on a case by case basis. The argument below is that the challenges themselves can be expected in high-dimensional cases, leading to the suggestion that they should be considered in the design of all multi-model forecast experiments.

The system used throughout this paper is the Moran-Ricker Map~\cite{Moran,Ricker} given in Equation~\ref{eq:MR}. Selection of a simple mathematical system allows the option of examining the challenges of a small forecast-outcome archive in the context of results based on huge archives. This is rarely possible for a physical system (see however~\cite{Machete08}). In this section the mathematical structure of the system and four imperfect models are specified. The specific structure of these models reflects a refined experimental design in light of the results of~\cite{Higgins15}.

%


Let $\tilde{x}_{t}$ be the state of Moran-Ricker Map at time $t\in\mathbb{Z}$. The evolution of the system state $\tilde{x}$ is given by the Moran-Ricker Map, i.e.
\begin{eqnarray}
 \label{eq:MR}
    \tilde{x}_{t+1}=\tilde{x}_{t}e^{\lambda(1-\tilde{x}_{t})}.
\end{eqnarray}
In the experiments presented in this paper, $\lambda=3$, where the system is somewhat ``less chaotic" than using the value adopted in {\cite{Sprott03}} (Figure 1 shows the Lyapunov exponent as a function of system parameter $\lambda$~\cite{Glendinning13}), in order to build models with comparable forecast skills. Define the observation at time $t$ to be $s_{t}=\tilde{x}_{t}+\eta_{t}$, where the observational noise, $\eta_{t}$, is independent normally distributed ($\eta_t \sim N(0, \sigma_{noise}^2)$)\footnote{Observations are restricted to positive values.}.

\begin{figure}[!h]

\hbox{
  \epsfig{file=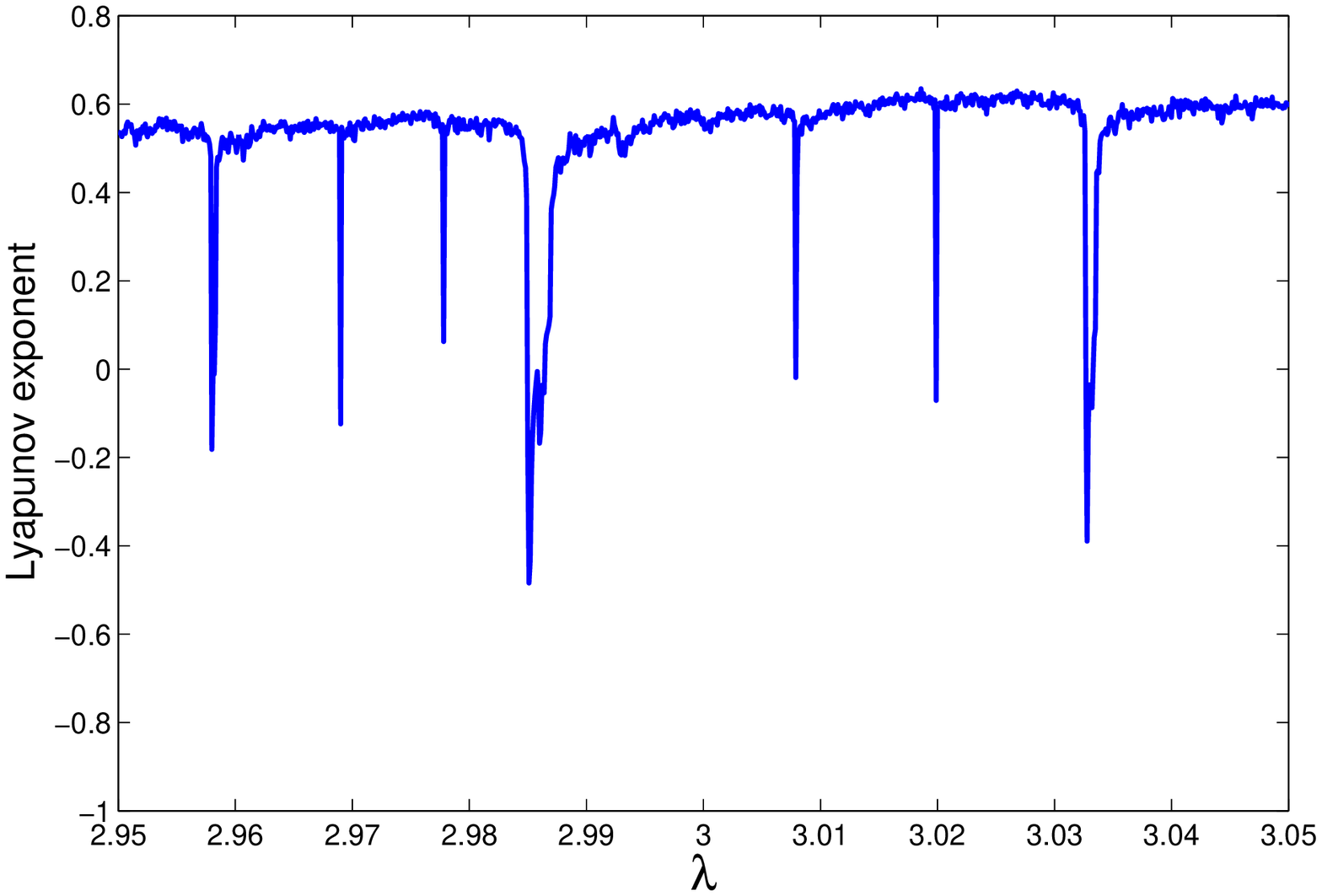, width=0.48\columnwidth, height=6cm}
  \epsfig{file=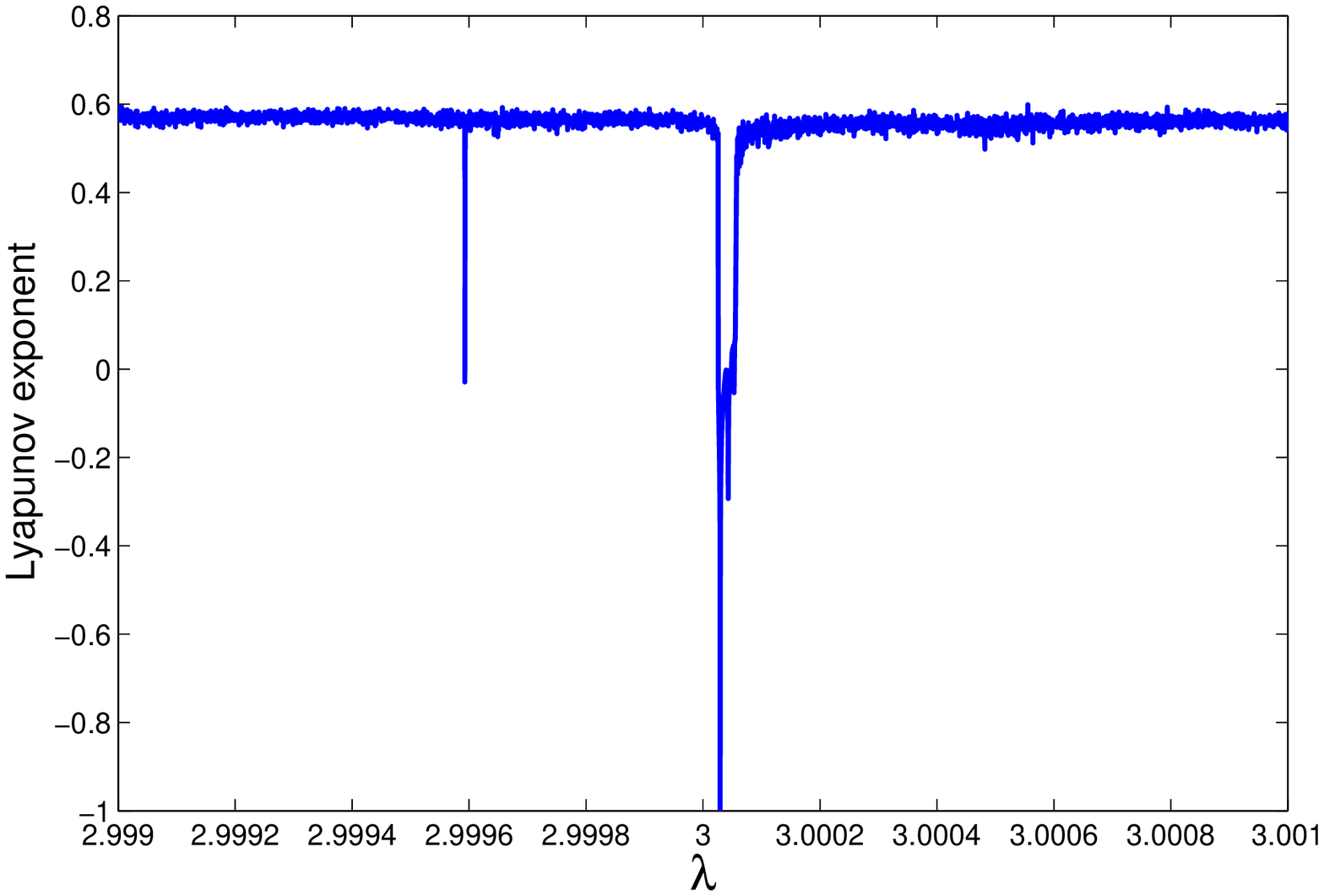, width=0.48\columnwidth, height=6cm}
}

\caption{Estimates of Global Lyapunov exponent are plotted as a function of $\lambda$. a) 4096 values of $\lambda$ uniformly random sampled between 2.95 and 3.05; b) 4096 values of $\lambda$ uniformly random sampled between 2.999 and 3.001.}
\label{fig:maperr1}
\end{figure}


Four one-dimensional deterministic models are constructed as imperfect models of the Moran-Ricker system. In experiments presented here, the focus is on designing ensemble scheme and ensemble parameter selection for producing predictive distribution, therefore the imperfect models as well as their parameter values are fixed. These four models share the same state space as the system, and the observations are complete. Note in practice, it is almost always the case that the model state $x$ lies in a different space from the system state $\tilde{x}$. The models are:

\begin{itemize}
\item \textbf{Model I}, $G_{1}(x)$, is built by first expanding the exponential term in Equation~\ref{eq:MR} to $12^{th}$ order:
\begin{eqnarray}
 \label{eq:M1_t}
    x_{t+1}=x(1+3(1-x)+\frac{1}{2!}(3(1-x))^{2}+\cdot\cdot\cdot+\frac{1}{12!}(3(1-x))^{12}).
\end{eqnarray}
The coefficient of each polynomial term is then truncated at the $3^{rd}$ decimal place:
\begin{eqnarray}
 \label{eq:M1}
    x_{t+1}=x(1+3(1-x)+4.5(1-x)^{2}+\cdot\cdot\cdot +0.004(1-x)^{11} +0.001(1-x)^{12}).
\end{eqnarray}

\item \textbf{Model II}, $G_{2}(x)$, is derived by first taking the log of Equation~\ref{eq:MR} and expanding to the $8^{th}$ order:
\begin{eqnarray}
 \label{eq:M2_t}
   log x_{t+1}=log x+3-3x = log x + 3 - 3 e ^ {log x}  \\
   log x_{t+1}=-2log x-\frac{3}{2!}(log x)^{2}-\frac{3}{3!}(log x)^{3}-\cdot\cdot\cdot-\frac{3}{8!}(log x)^{8}
\end{eqnarray}
The coefficient of each polynomial term is then truncated at the $4^{th}$ decimal place:
\begin{eqnarray}
 \label{eq:M2}
   log x_{t+1}=-2log x-1.5(log x)^{2}-0.5(log x)^{3}-\cdot\cdot\cdot -0.0006(log x)^{7} -0.0001(log x)^{8}
\end{eqnarray}

\item \textbf{Model III}, $G_{3}(x)$, is obtained by expanding the right hand side of Equation~\ref{eq:MR} in a Fourier series over the range $0\leq \tilde{x} \leq \pi$. This series is then truncated at the $10^{th}$ order to yield

\begin{eqnarray}
 \label{eq:M3}
   x_{t+1}=\frac{a_{0}}{2}+\sum^{10}_{i=1}[a_{i}cos(2 i x_{t})+b_{i}sin(2 i x_{t})], \nonumber
\end{eqnarray}
where the coefficients $a_{i}$ and $b_{i}$ are obtained by
\begin{eqnarray}
 \label{eq:M3coef}
   a_{i}=\frac{2}{\pi}\int^{\pi}_{0}xe^{\lambda(1-x)}cos(2ix)dx \\
   b_{i}=\frac{2}{\pi}\int^{\pi}_{0}xe^{\lambda(1-x)}sin(2ix)dx
\end{eqnarray}

\item \textbf{Model IV}, $G_{4}(x)$, is defined by

\begin{eqnarray}
 \label{eq:M4}
   x_{t+1}=x_{t}e^{\lambda(1-\delta-x_{t})},
\end{eqnarray}
where $\delta$ is taken to be 0.02. Technically, the addition of $\delta$ means model IV is a case of parameter uncertainty, {\bf NOT} structural model error. 

\end{itemize}


Notice that the order of the truncation for model I,II and III are different. The order of the truncation for the first three models and also the parameter $\delta$ for model IV are chosen so that those models present the system dynamics well and the scales of their forecast skill are comparable. Figure~\ref{fig:maperr1} plots the model dynamics of each model together with the system dynamics. Figure~\ref{fig:histerr1} presents the histogram of the 1-step model error over 2048 different initial conditions. It appears that Model I simulates the system dynamics well except when the initial condition is near the maximum value of the system. For Model II, the difference between model dynamics and system dynamics appears only when the initial condition is near the minimum value of the system. Model III doesn't match the system dynamics well where $x\gtrsim 1.5$ and where $F(x)$ reaches the maximum value of the map. Model IV mapping the initial condition to lower values comparing with the system, the larger the images are, the more differences. 

Figure~\ref{fig:maperr2} plots the two-step model error for each model, while Figure~\ref{fig:histerr2} presents the histogram of the 2-step model error. Generally the structure of the model error is different. Different models have different scales of model error in different state space. 

\begin{figure}[!h]
\hbox{
  \epsfig{file=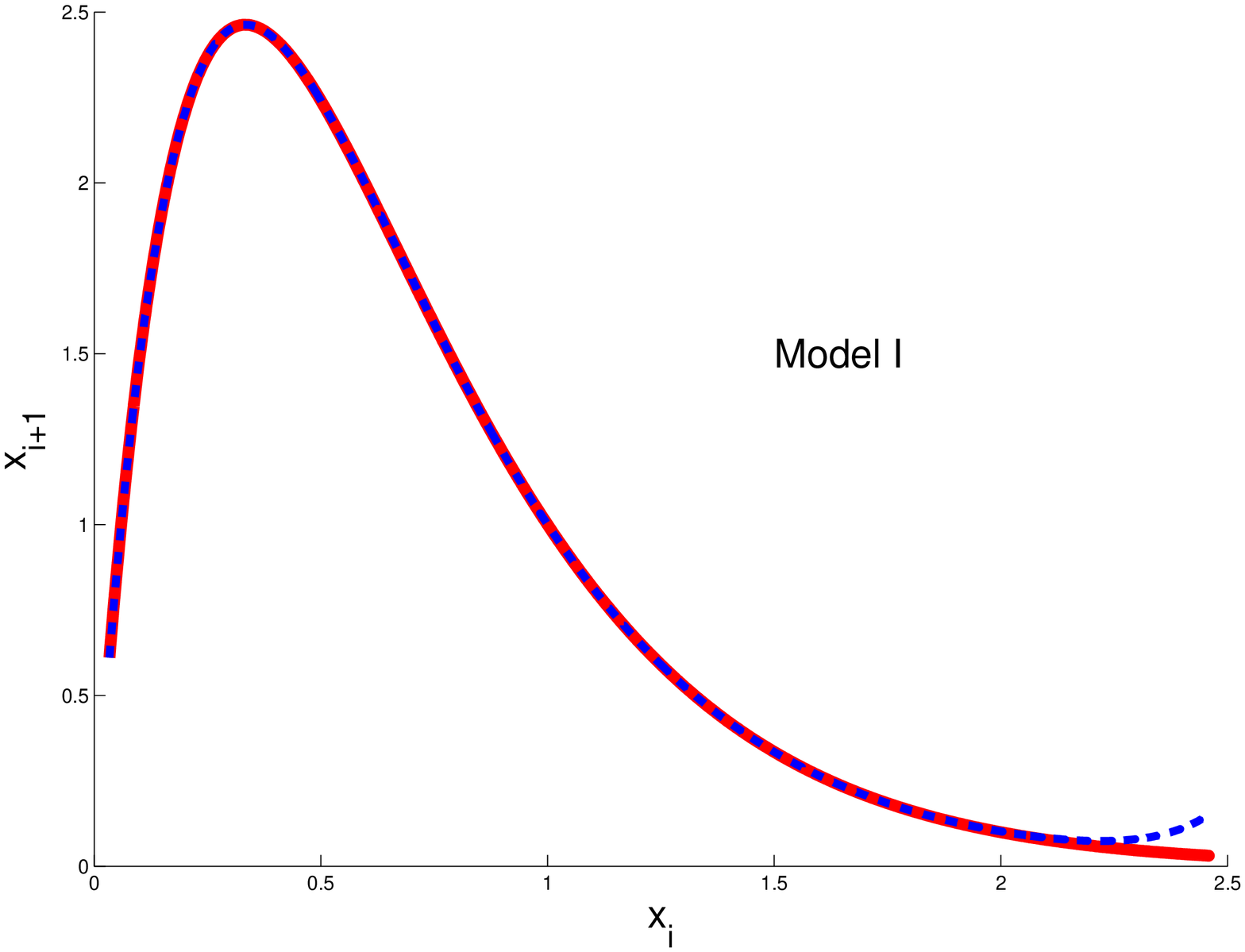, width=0.48\columnwidth, height=6cm}
  \epsfig{file=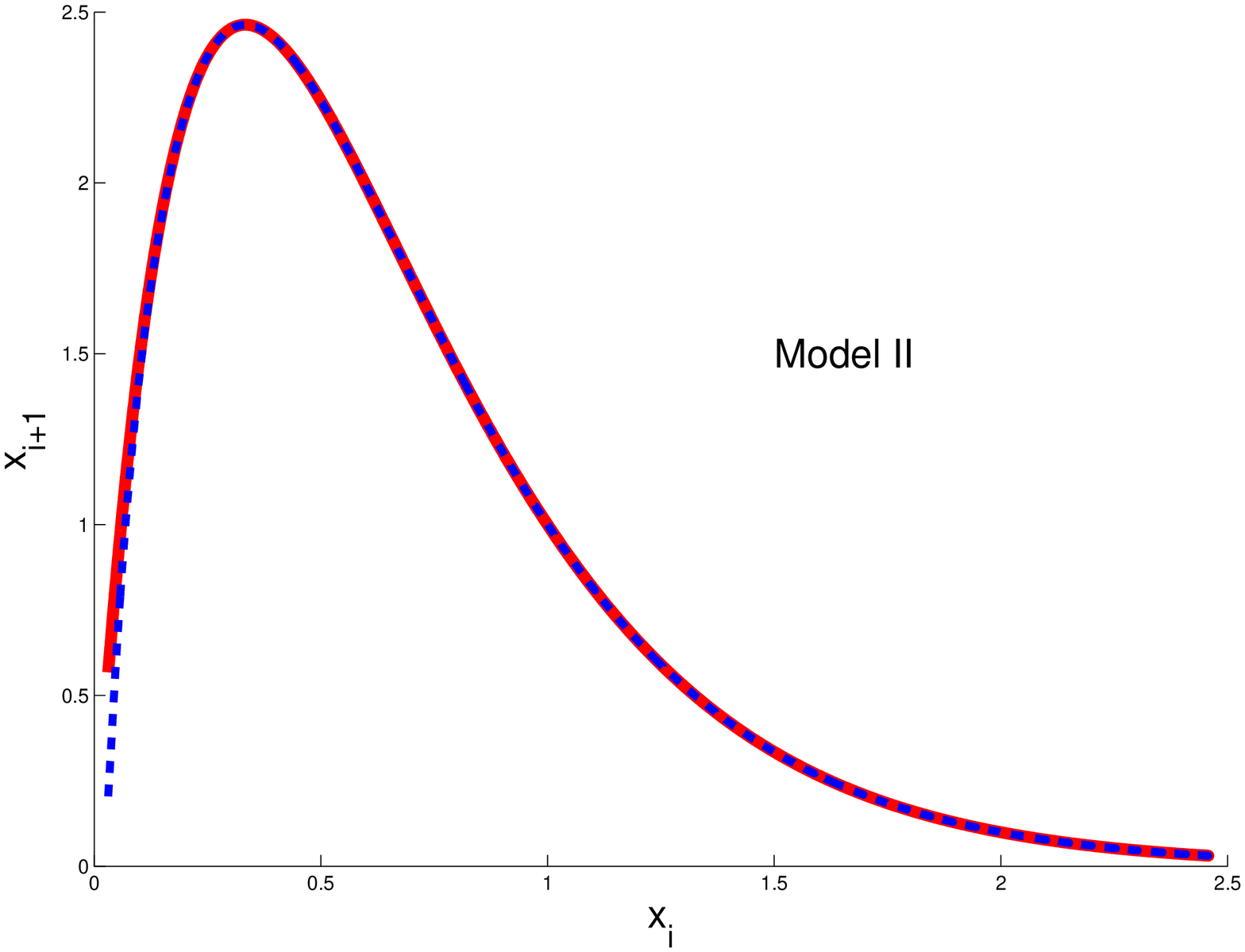, width=0.48\columnwidth, height=6cm}
}
\hbox{
  \epsfig{file=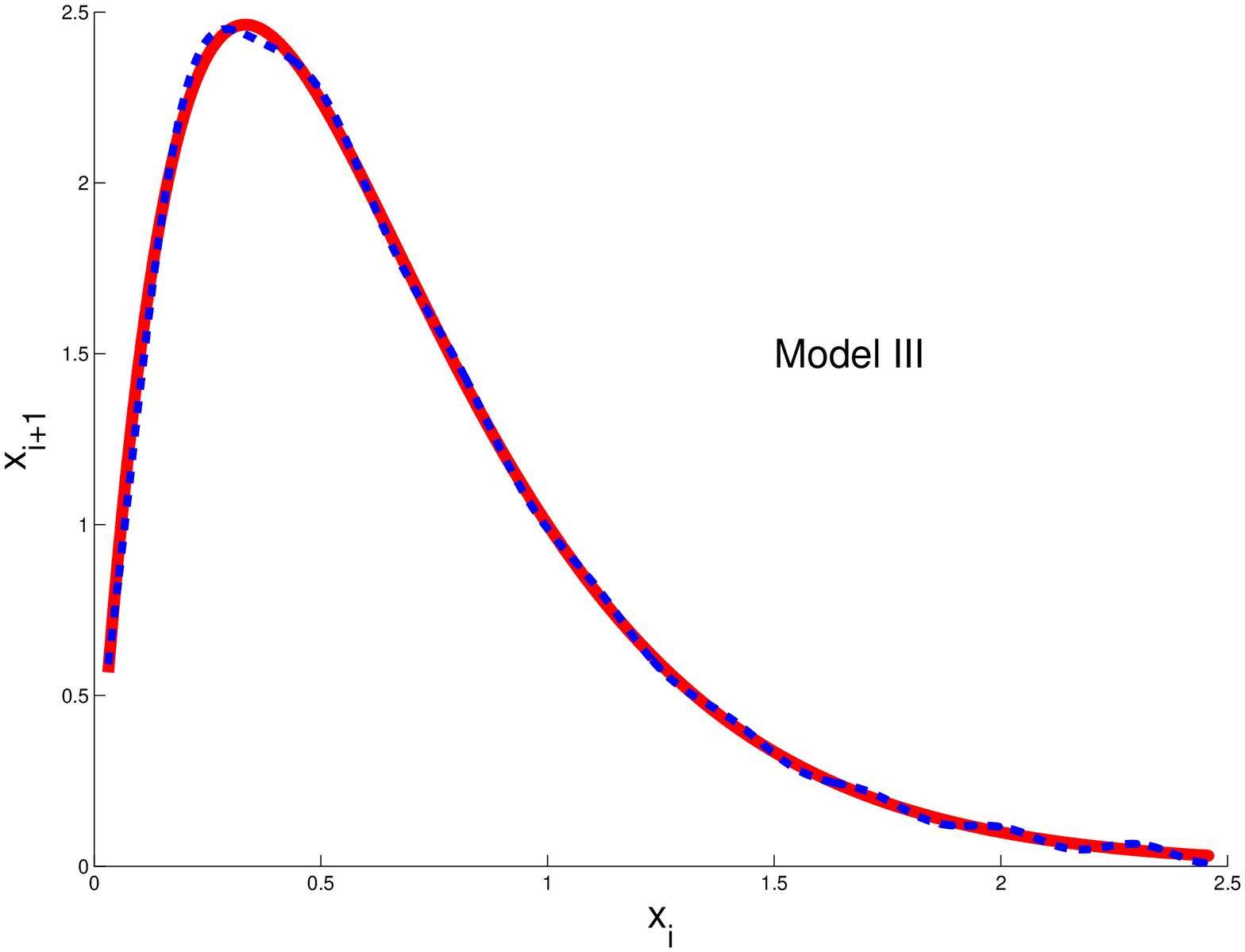, width=0.48\columnwidth, height=6cm}
  \epsfig{file=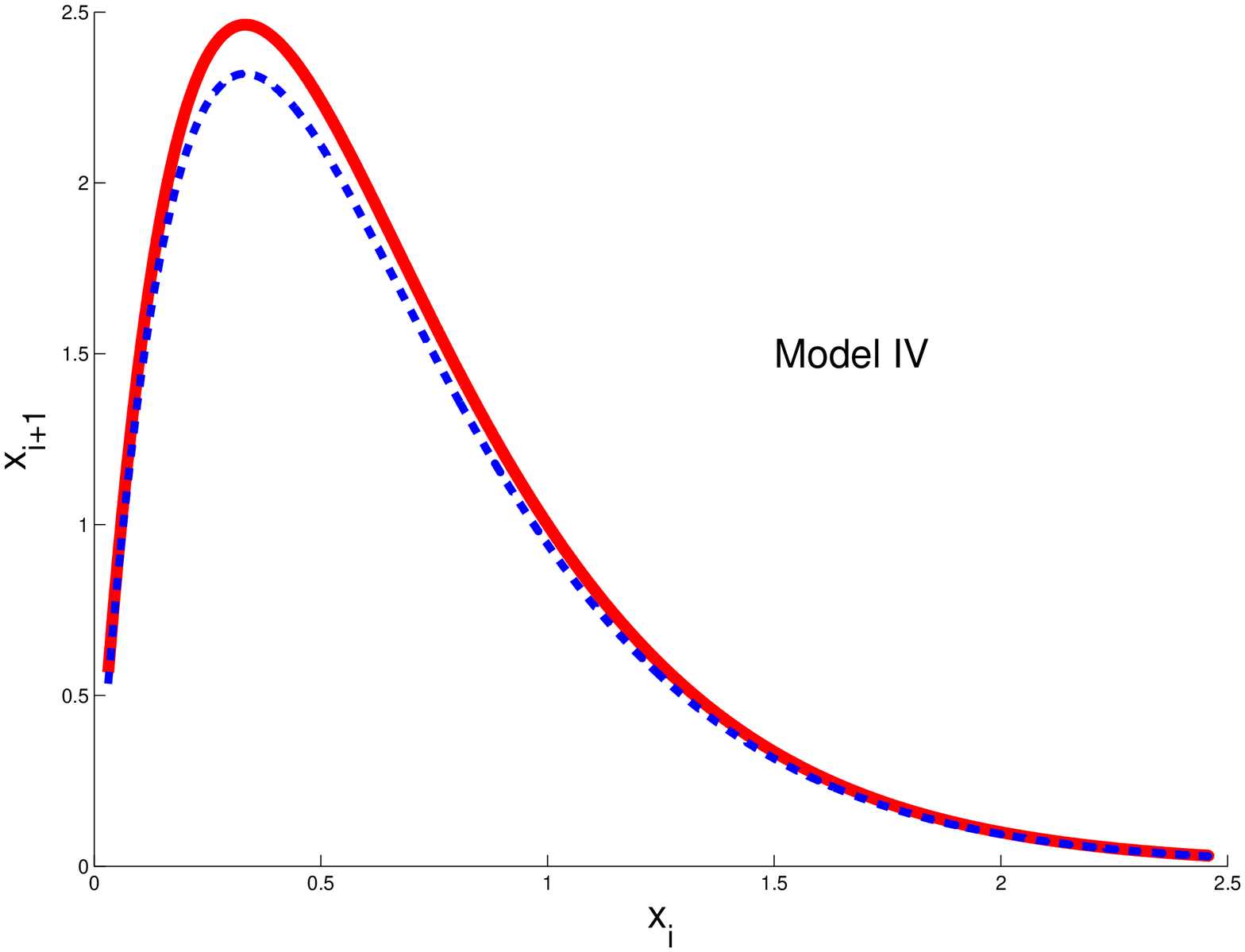, width=0.48\columnwidth, height=6cm}
}
\caption{Graphical presentation of the dynamics of four different models, the blue line represents model dynamics as a function of initial conditions and the red line represents the system dynamics.}
\label{fig:maperr1}
\end{figure}

\begin{figure}[!h]
\hbox{
  \epsfig{file=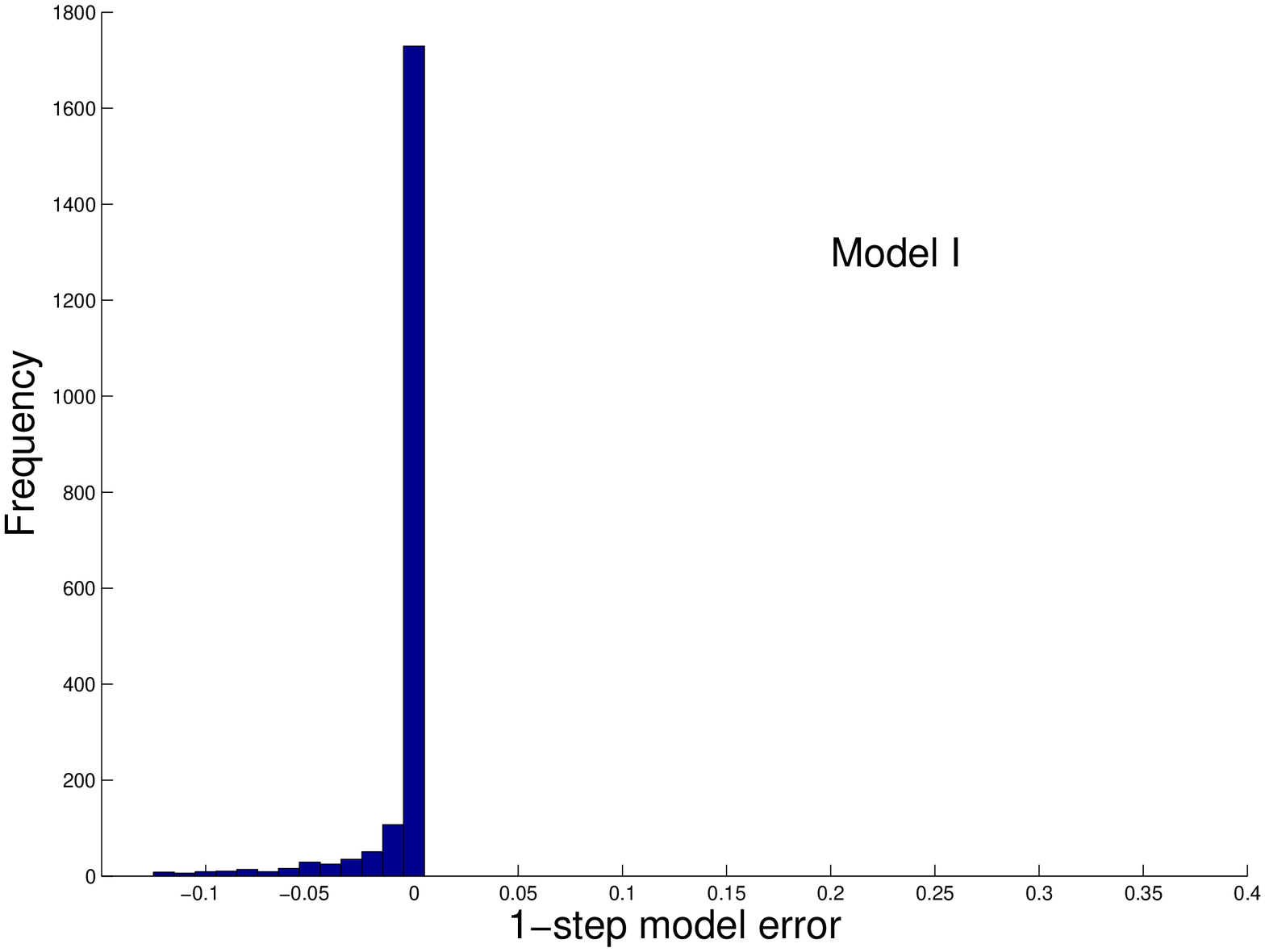, width=0.48\columnwidth, height=6cm}
  \epsfig{file=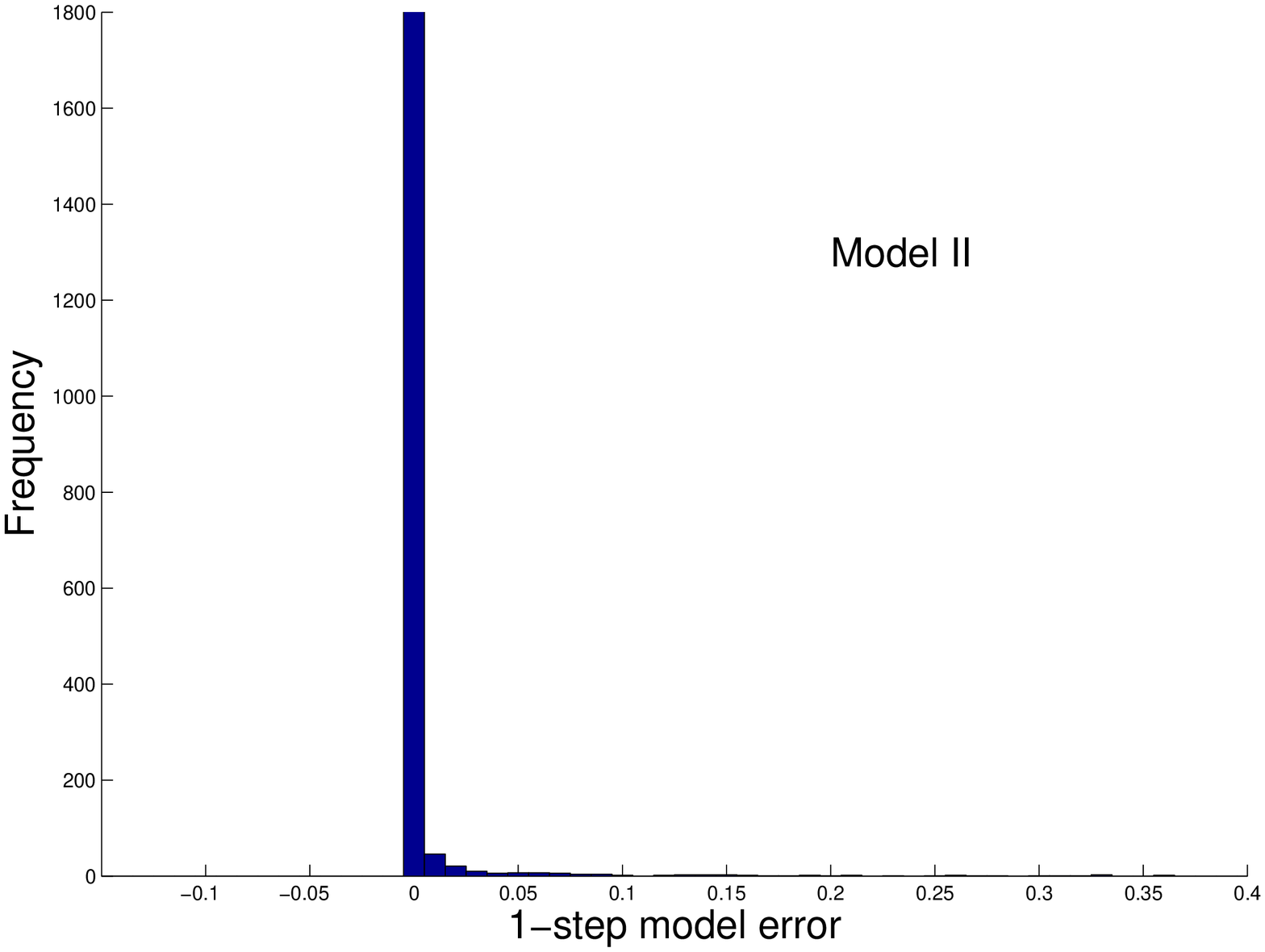, width=0.48\columnwidth, height=6cm}
}
\hbox{
  \epsfig{file=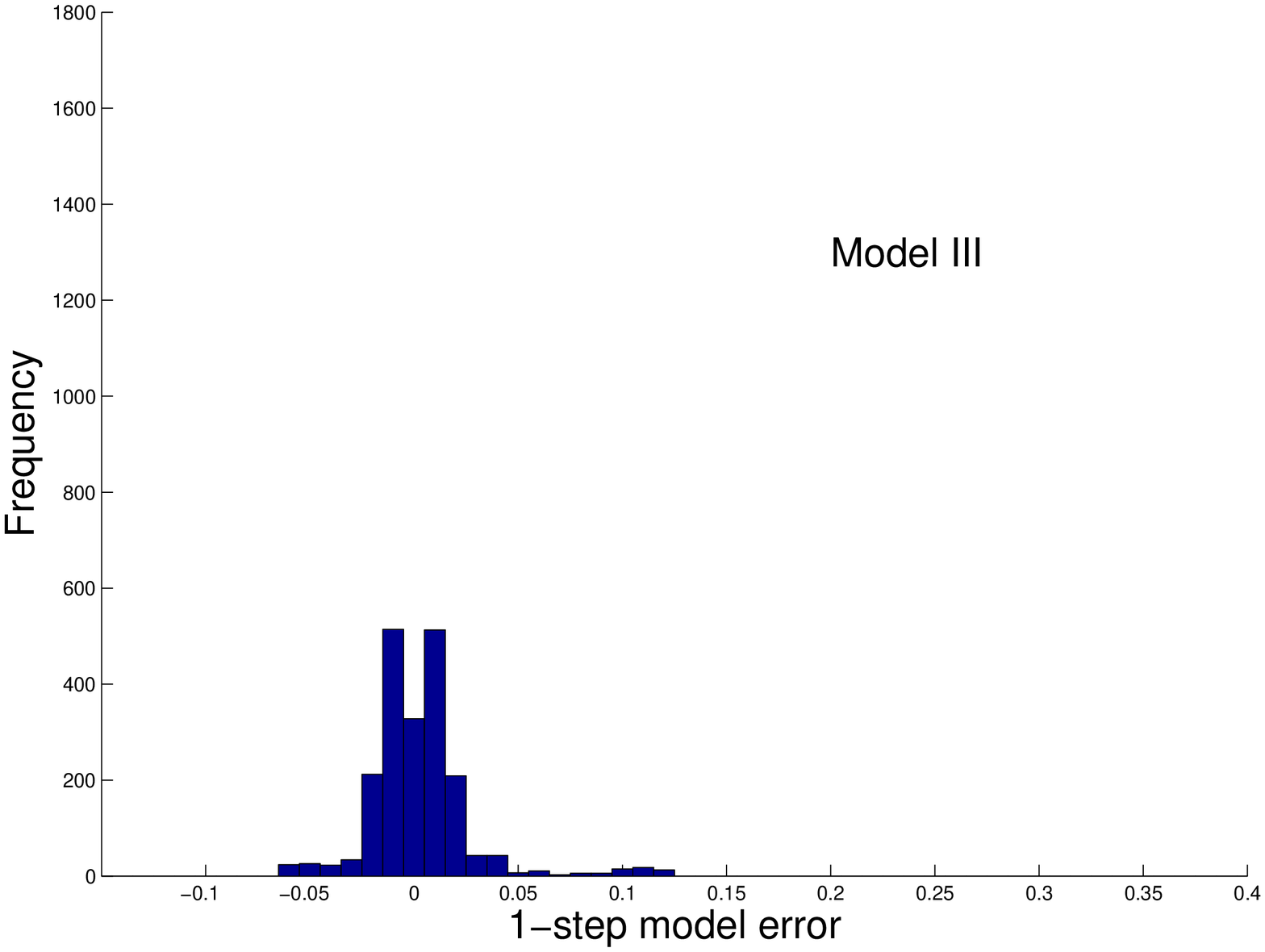, width=0.48\columnwidth, height=6cm}
  \epsfig{file=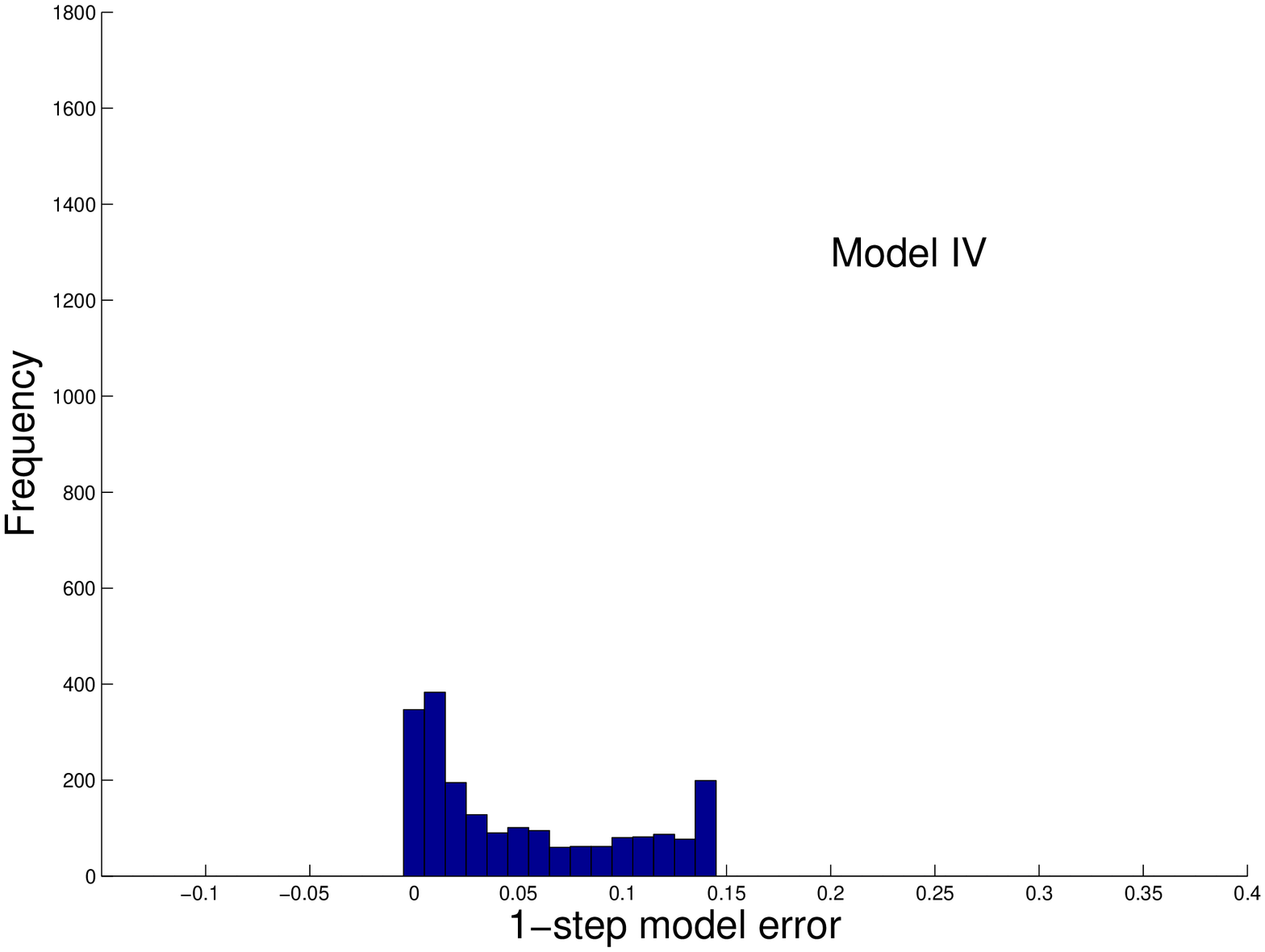, width=0.48\columnwidth, height=6cm}
}
\caption{Histogram of the 1-step model errors, given 2048 different initial conditions with respect to natural measure.}
\label{fig:histerr1}
\end{figure}

\begin{figure}[!h]
\hbox{
  \epsfig{file=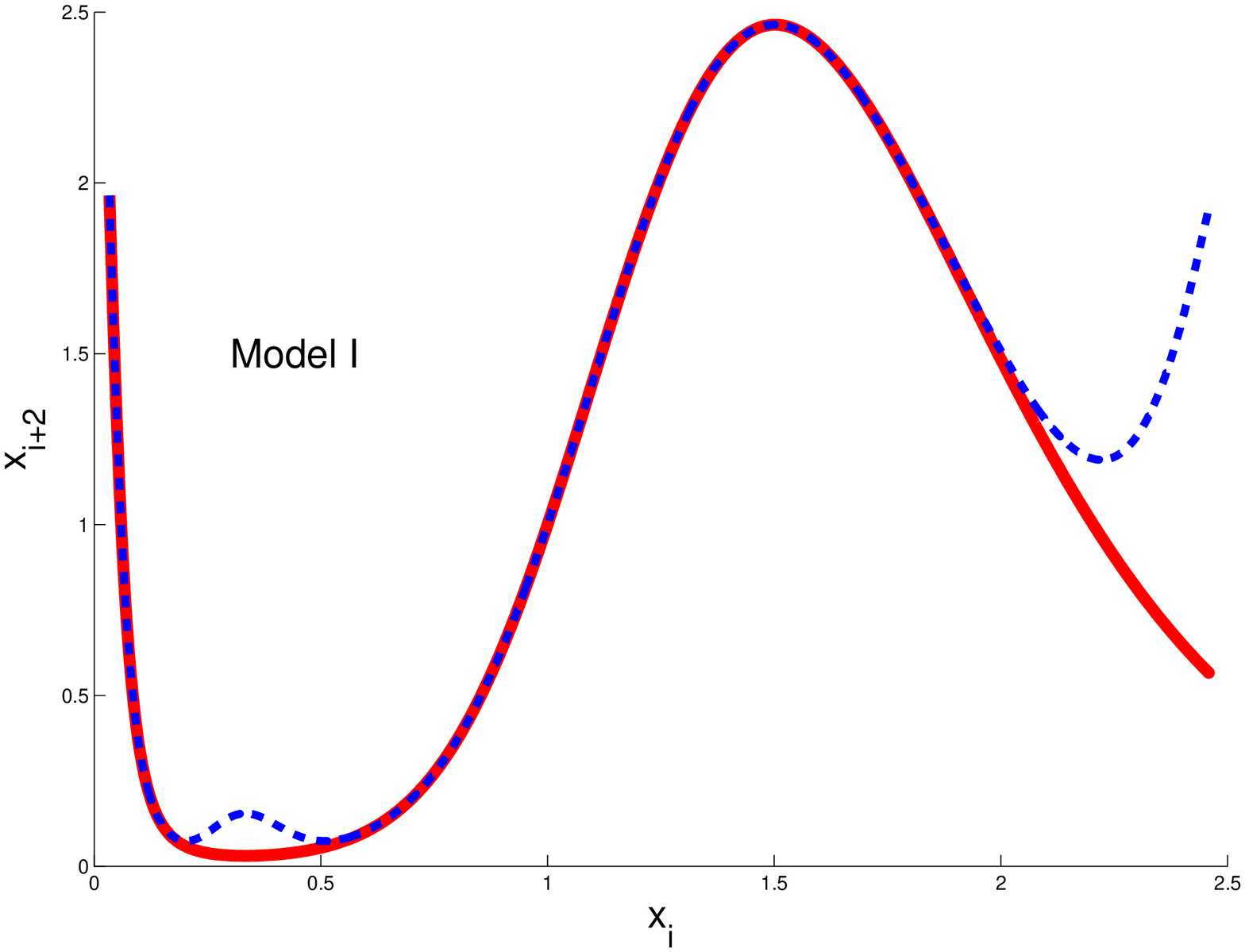, width=0.48\columnwidth, height=6cm}
  \epsfig{file=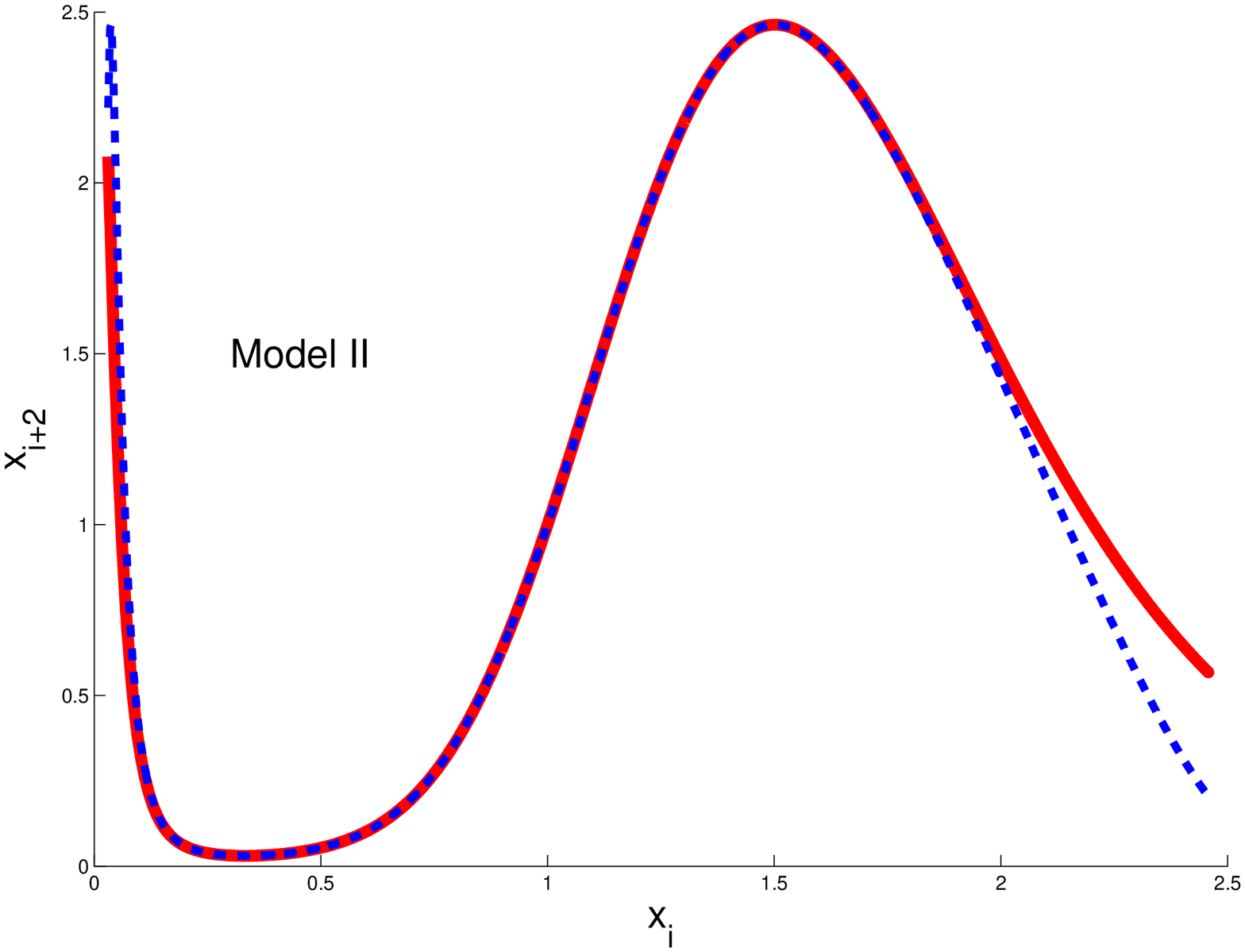, width=0.48\columnwidth, height=6cm}
}
\hbox{
  \epsfig{file=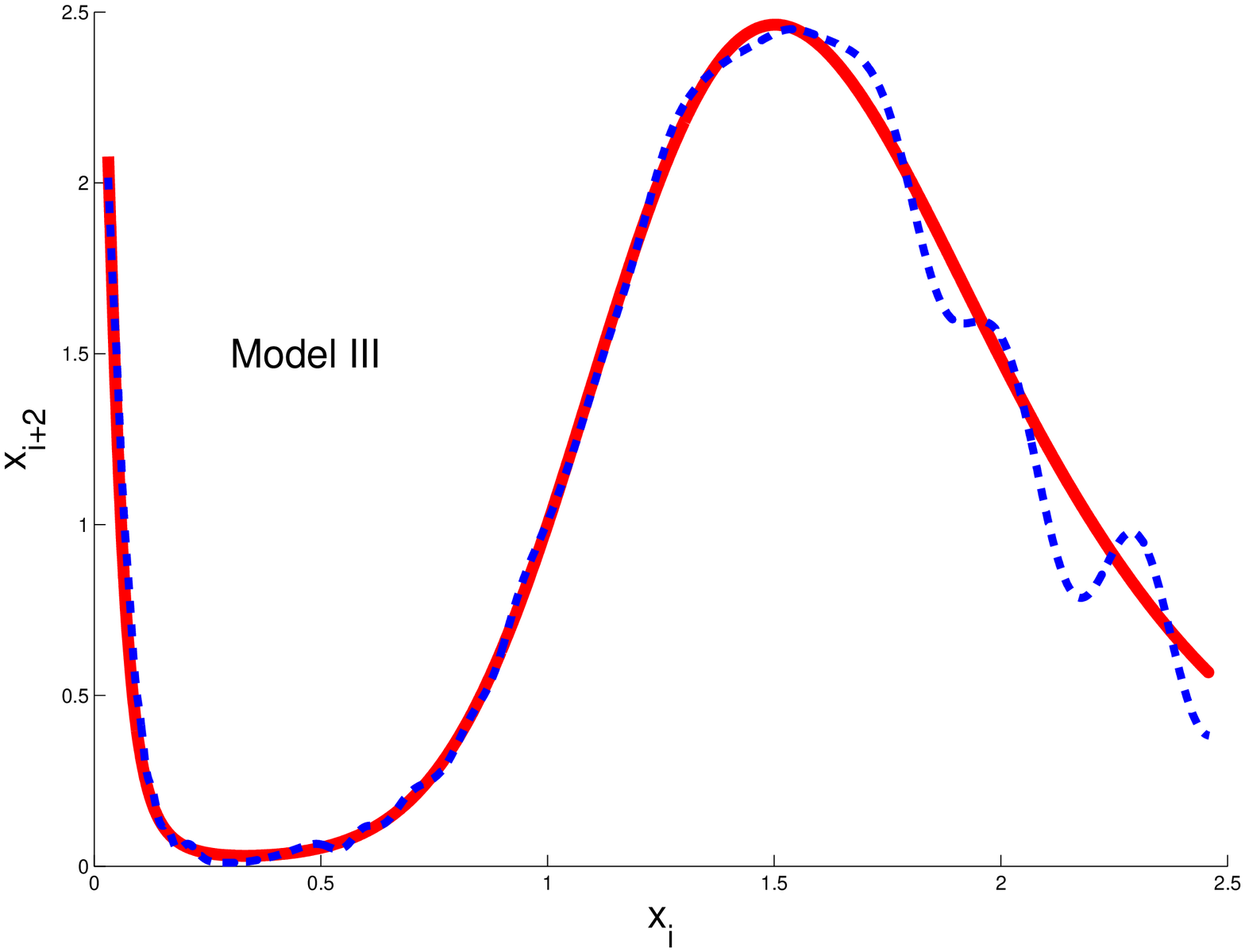, width=0.48\columnwidth, height=6cm}
  \epsfig{file=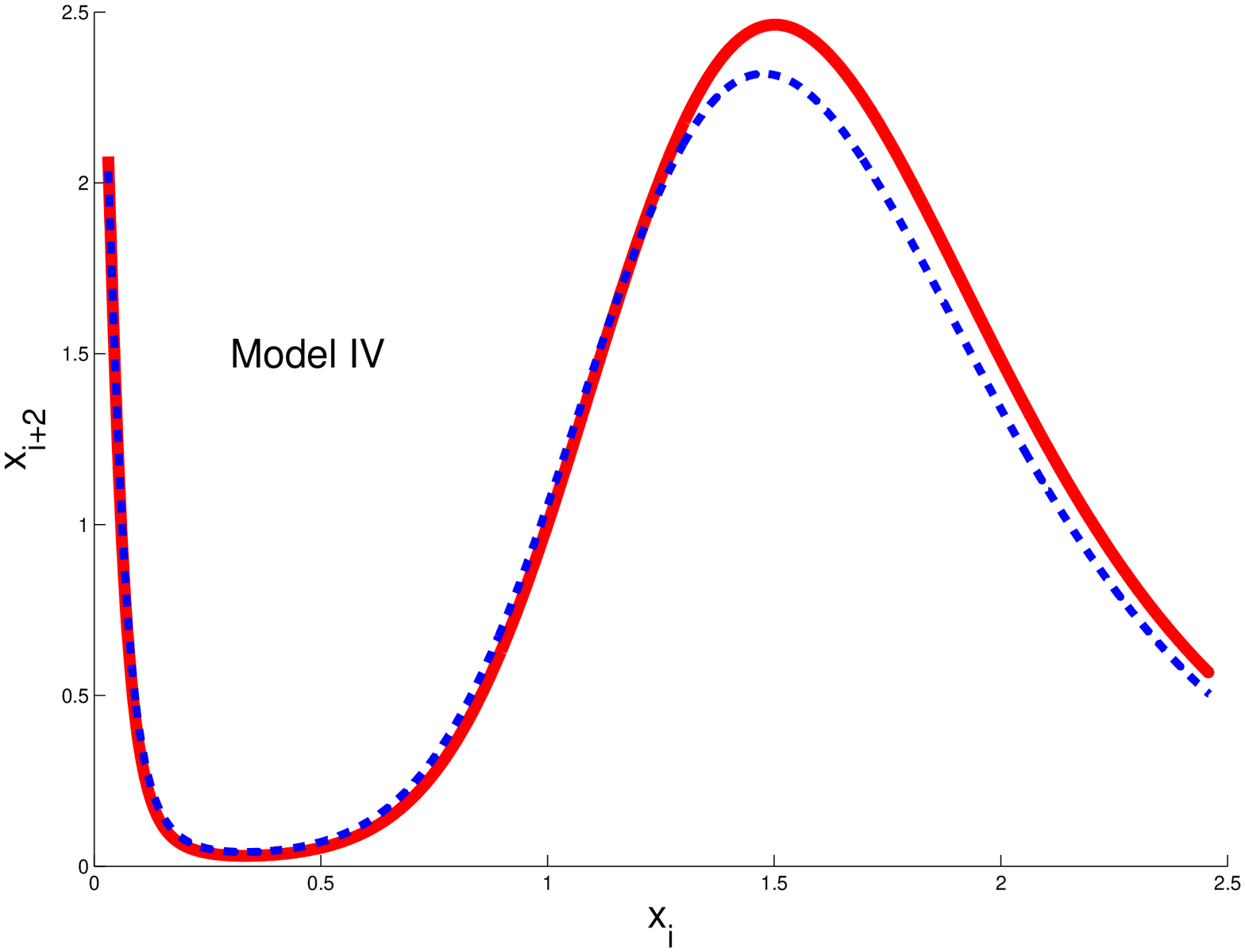, width=0.48\columnwidth, height=6cm}
}
\caption{Graphical presentation of the 2-step evolution of four different models, the blue line represents 2-step model evolution as a function of initial conditions and the red line represents the 2-step evolution under the system.}
\label{fig:maperr2}
\end{figure}

\begin{figure}[!h]
\hbox{
  \epsfig{file=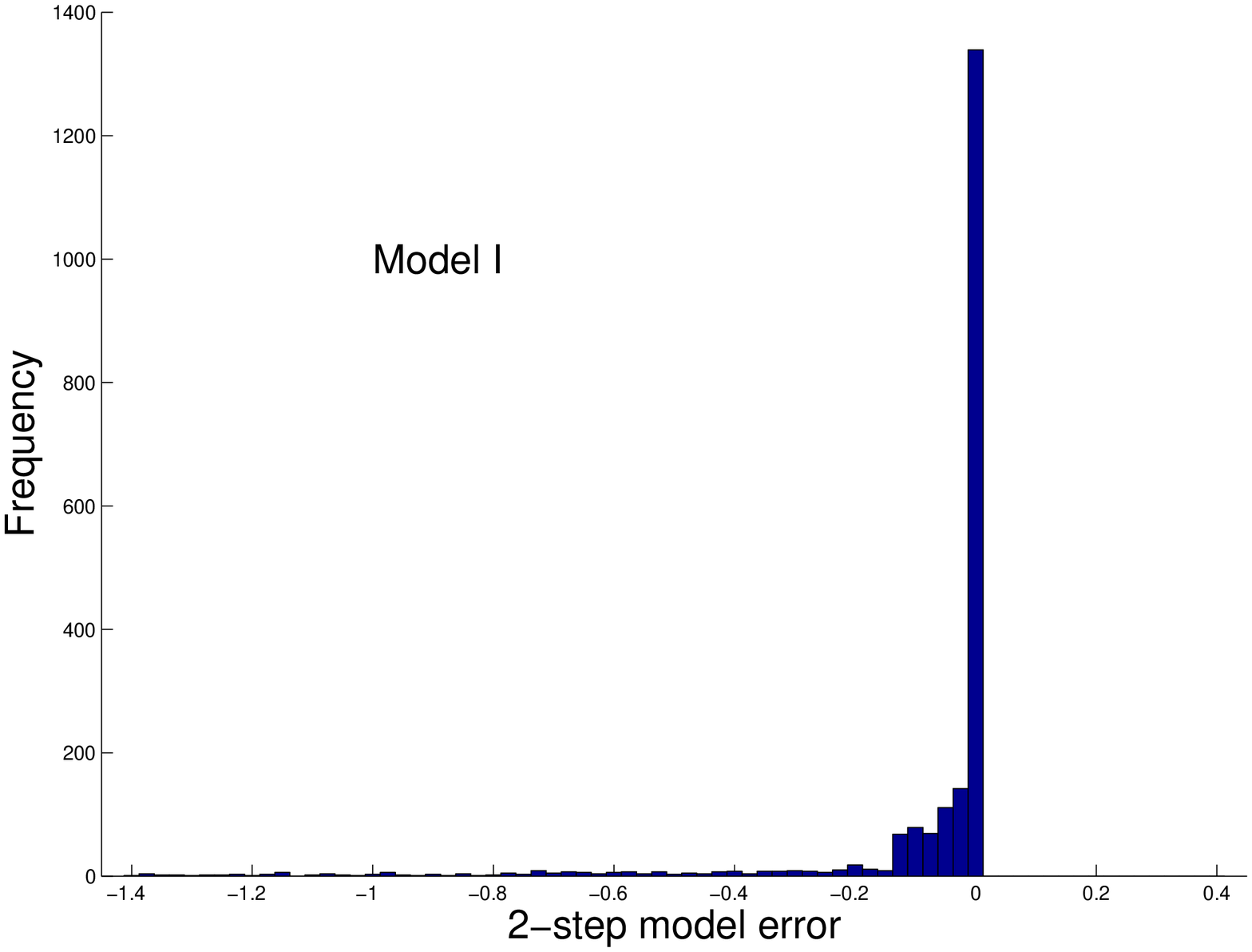, width=0.48\columnwidth, height=6cm}
  \epsfig{file=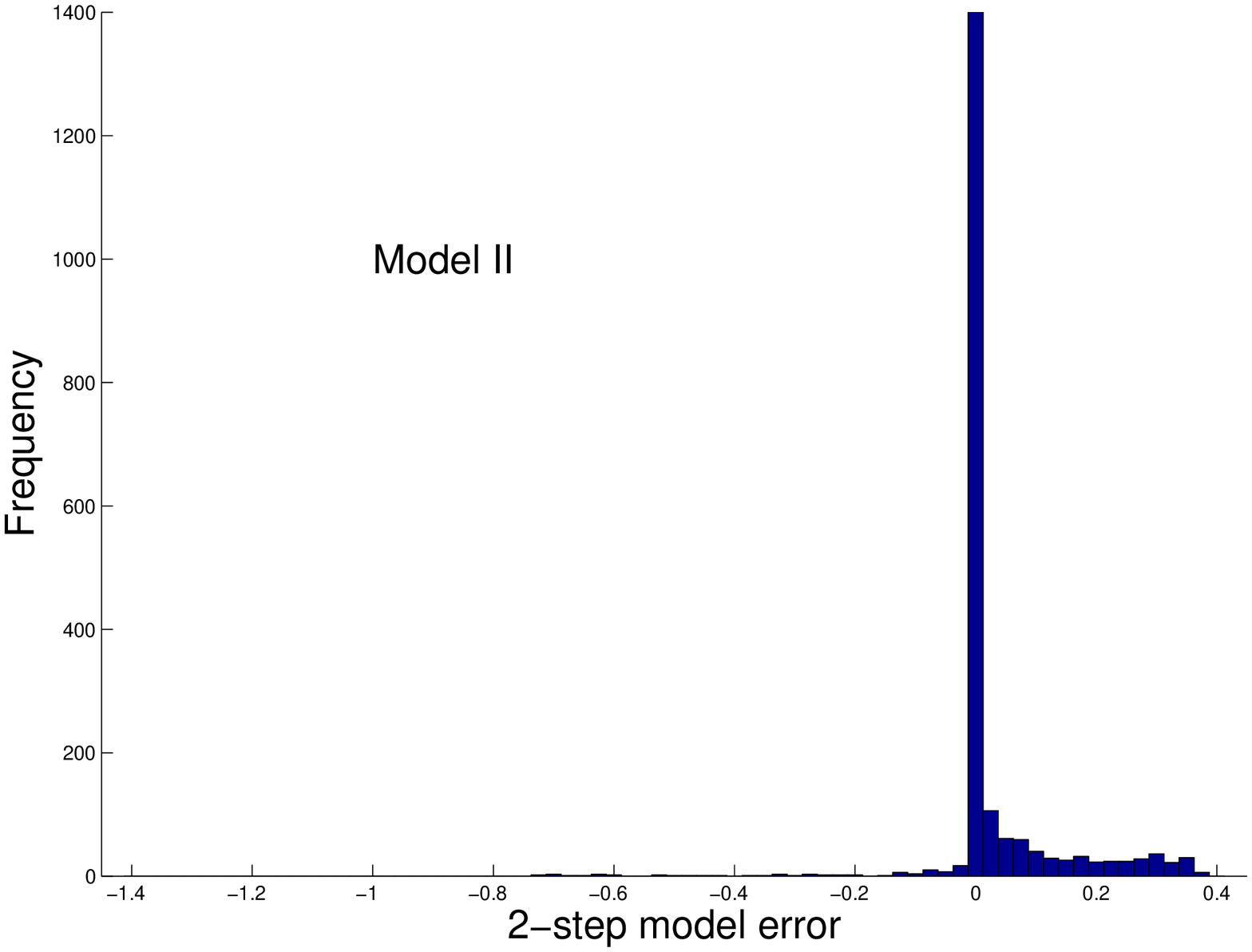, width=0.48\columnwidth, height=6cm}
}
\hbox{
  \epsfig{file=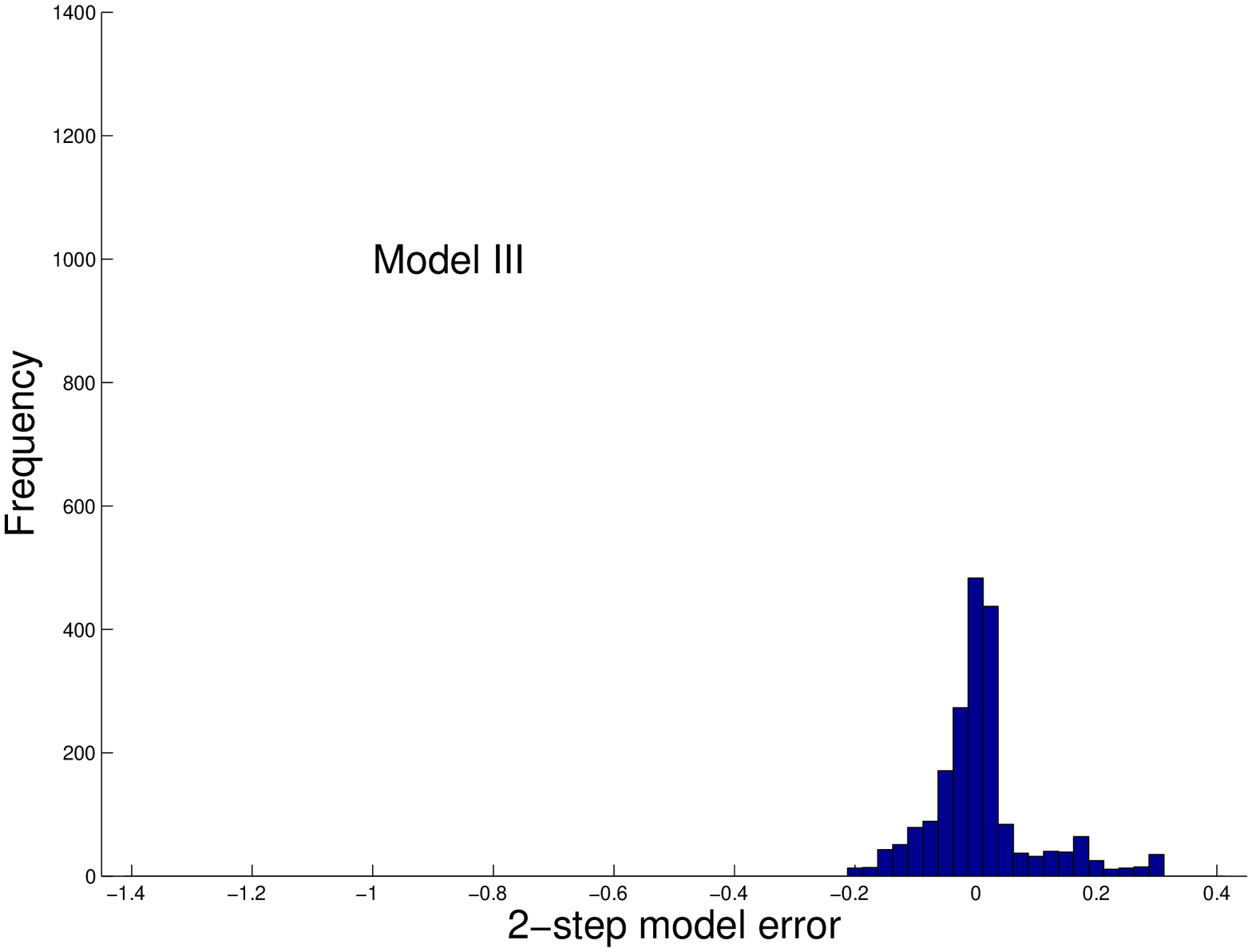, width=0.48\columnwidth, height=6cm}
  \epsfig{file=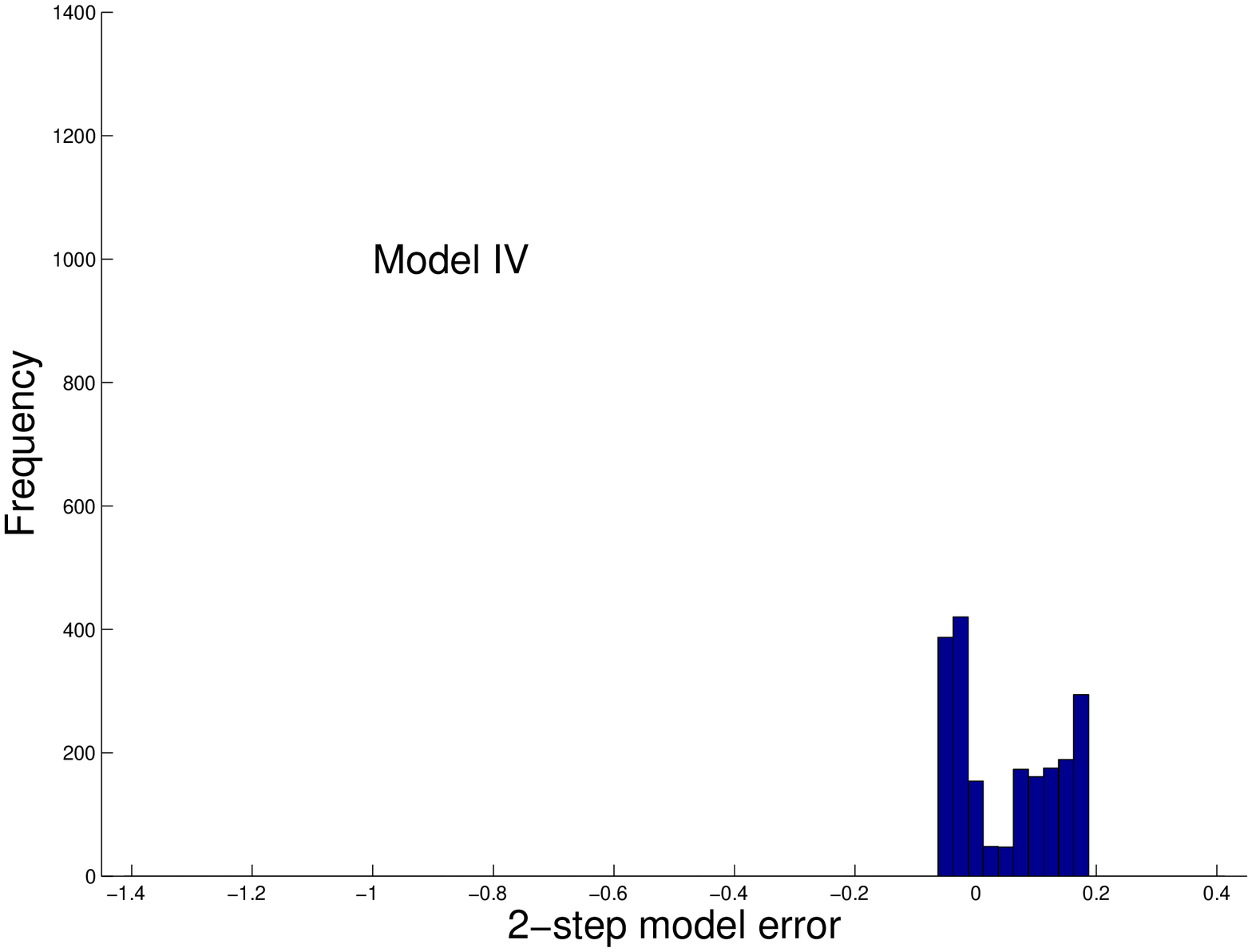, width=0.48\columnwidth, height=6cm}
}
\caption{Histogram of the 2-step model errors, given 2048 different initial conditions with respect to natural measure.}
\label{fig:histerr2}
\end{figure}

{Again, there is, of course, no suggestion that the Moran-Ricker system resembles the dynamics of the Earth. Rather, the framework presented here (and in~\cite{Higgins15}) provides probability forecasts from structurally flawed models; both the model-based forecasts (and an ideal probability forecast given the system) differ nontrivially from each other, and as the models are nonlinear the forecast distributions are non-Gaussian. It is these challenges to multi-model forecast system development which are illustrated in this paper, which should (of course) not be taken to present an actual geophysical forecast system; indeed the computational requirements and length of the observational record would arguably preclude examination of LAP of ``state of the art" geophysical models.}




\section{IC ensembles for each model}



In the experiments presented in this paper, each model produces its ensemble forecasts by iterating an ensemble of initial conditions (IC). The initial condition ensemble is formed by perturbing the observation with random draws from a Normal distribution, $N(0, \kappa^2_{\tau})$. The perturbation parameter $\kappa_{\tau}$ is chosen to minimize the Ignorance score at lead time $\tau$. When making medium-range forecasts, ECMWF selects a perturbation size such that the RMS error between the ensemble members and ensemble mean at a lead time of two days is roughly equal to the RMS of ensemble mean and the outcome at two days.

\begin{figure}[!h]
\hbox{
  \epsfig{file=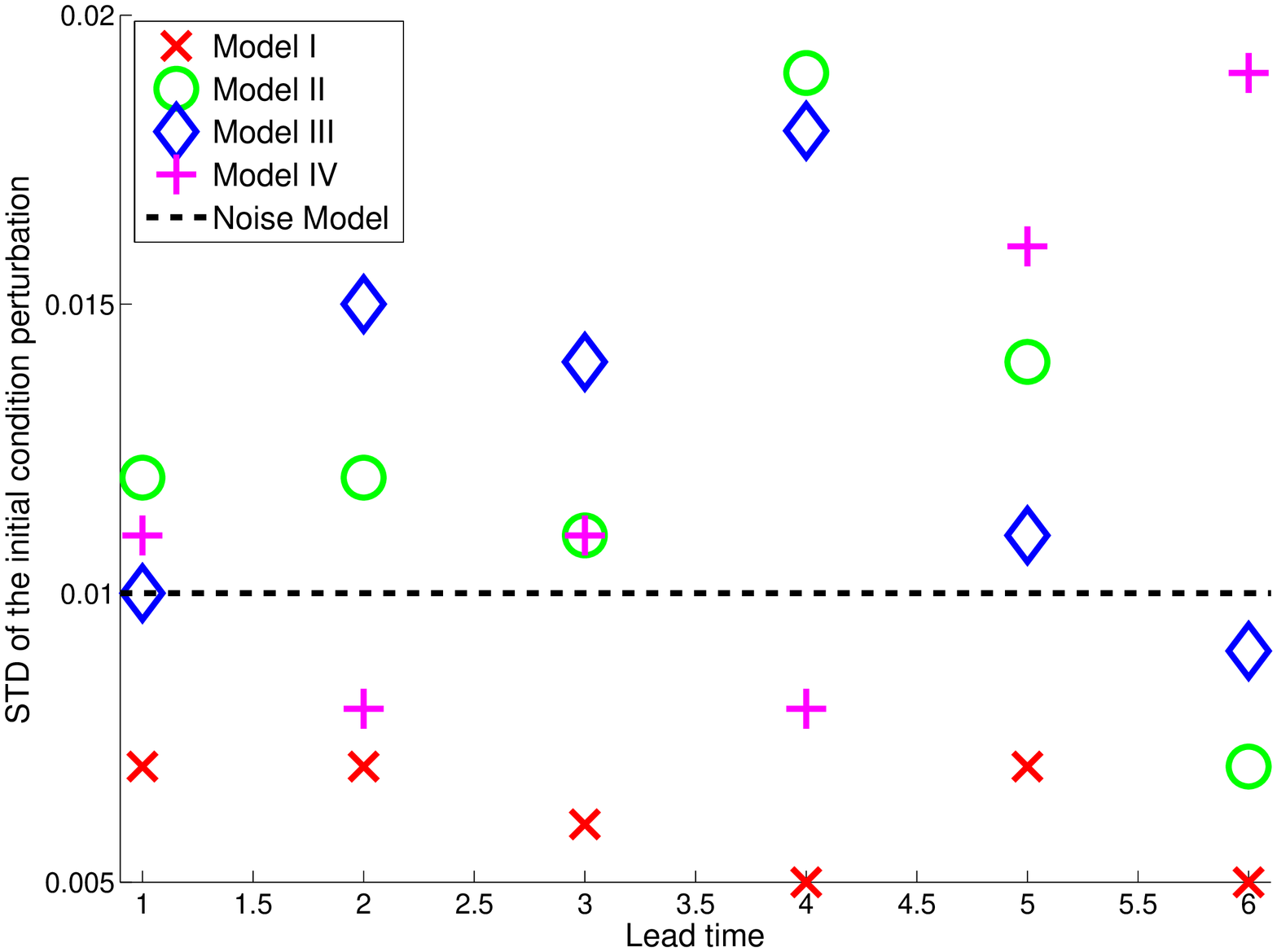, width=0.48\columnwidth, height=6cm}
  \epsfig{file=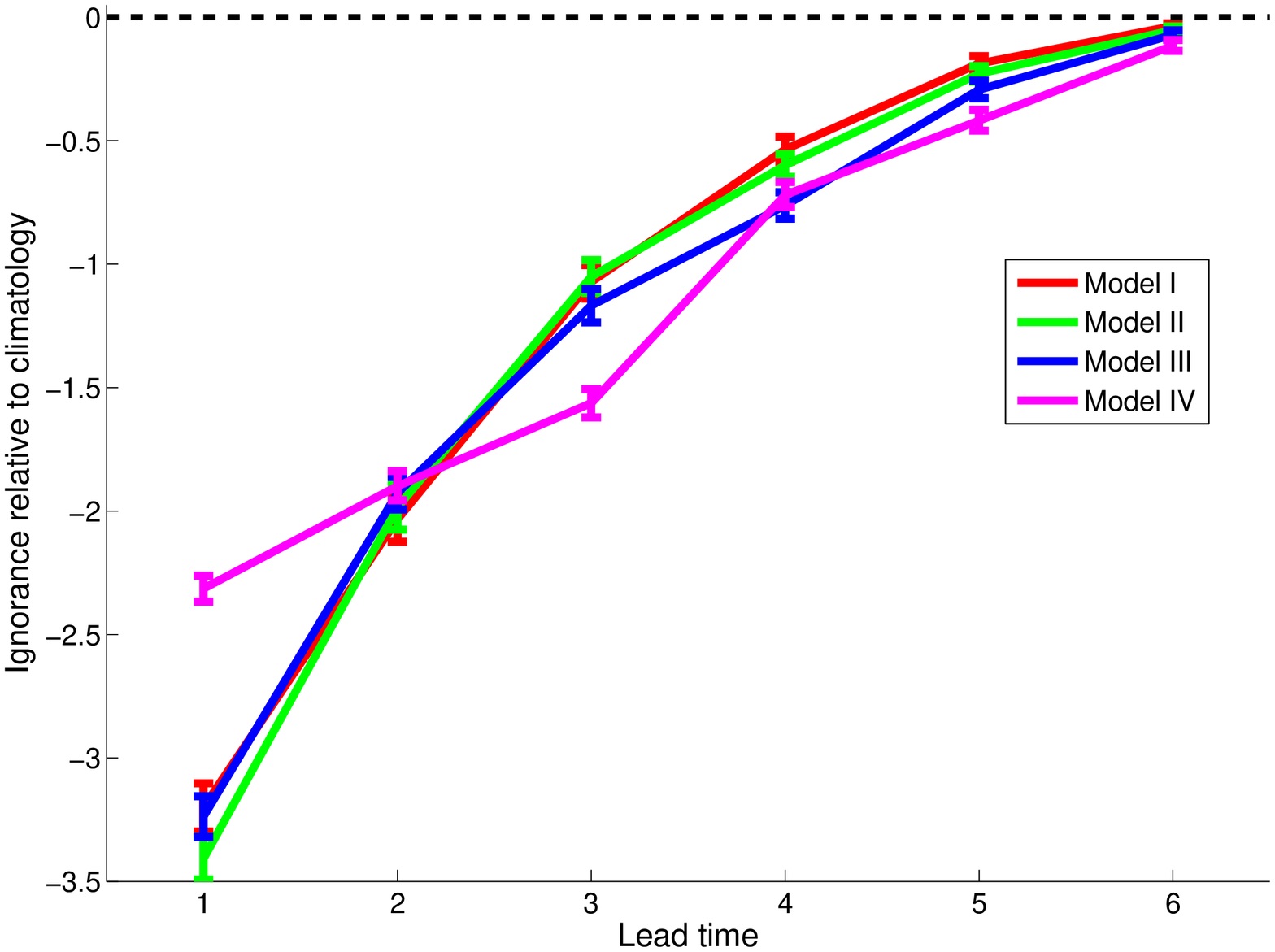, width=0.48\columnwidth, height=6cm}
}

\caption{a) The best found perturbation parameter values $\kappa$ as a function of lead time for each model, the dashed black line reflects the standard deviation of the noise model. b) Ignorance score of each model as a function of lead time, the dashed black line reflects the zero skill climatology. }
\label{fig:kappa}
\end{figure}

In experiments presented here, each initial condition ensemble will contain $N_{e}= 9$ members, following the ENSEMBLES protocol. Consider first the case of a large archive, with $N_{a} = 2048$. 
For a given $\kappa$ and lead time $\tau$, the kernel dressing and climatology-blend parameter values are fitted using a training forecast-outcome set which contains 2048 forecast-outcome pairs. The Ignorance score is then calculated using an independent testing forecast-outcome set which also contains 2048 forecast-outcome pairs. Figure~\ref{fig:kappa}a shows the best found perturbation parameter $\kappa$ for each model varies with lead time. The Ignorance score for each model at different lead time, using the values of $\kappa$ in Figure~\ref{fig:kappa} a, is shown in Figure~\ref{fig:kappa} b. As seen in figure 6a, for each model the preferred value of $\kappa$ varies significantly (about a factor of 2) between different lead times. Defining a $N_{e}$ member forecast system requires selecting a single value of $\kappa$ for each model. In this paper, the value of $\kappa$ for each model is chosen by optimizing the forecast Ignorance score at lead time 1. Sensitivity tests have been conducted and the Ignorance score at other lead times is much less sensitive to $\kappa$ than that at lead time 1.

\section{On the number of IC simulations in each Ensemble}

Knowledge of the relationship between ensemble size and forecast quality aids forecast system design. The cost of increasing the number of ensemble members is typically small relative to the cost of developing a new model. The cost of increasing the ensemble size increases only (nearly) linearly. 

\begin{figure}[!h]
\hbox{
  \epsfig{file=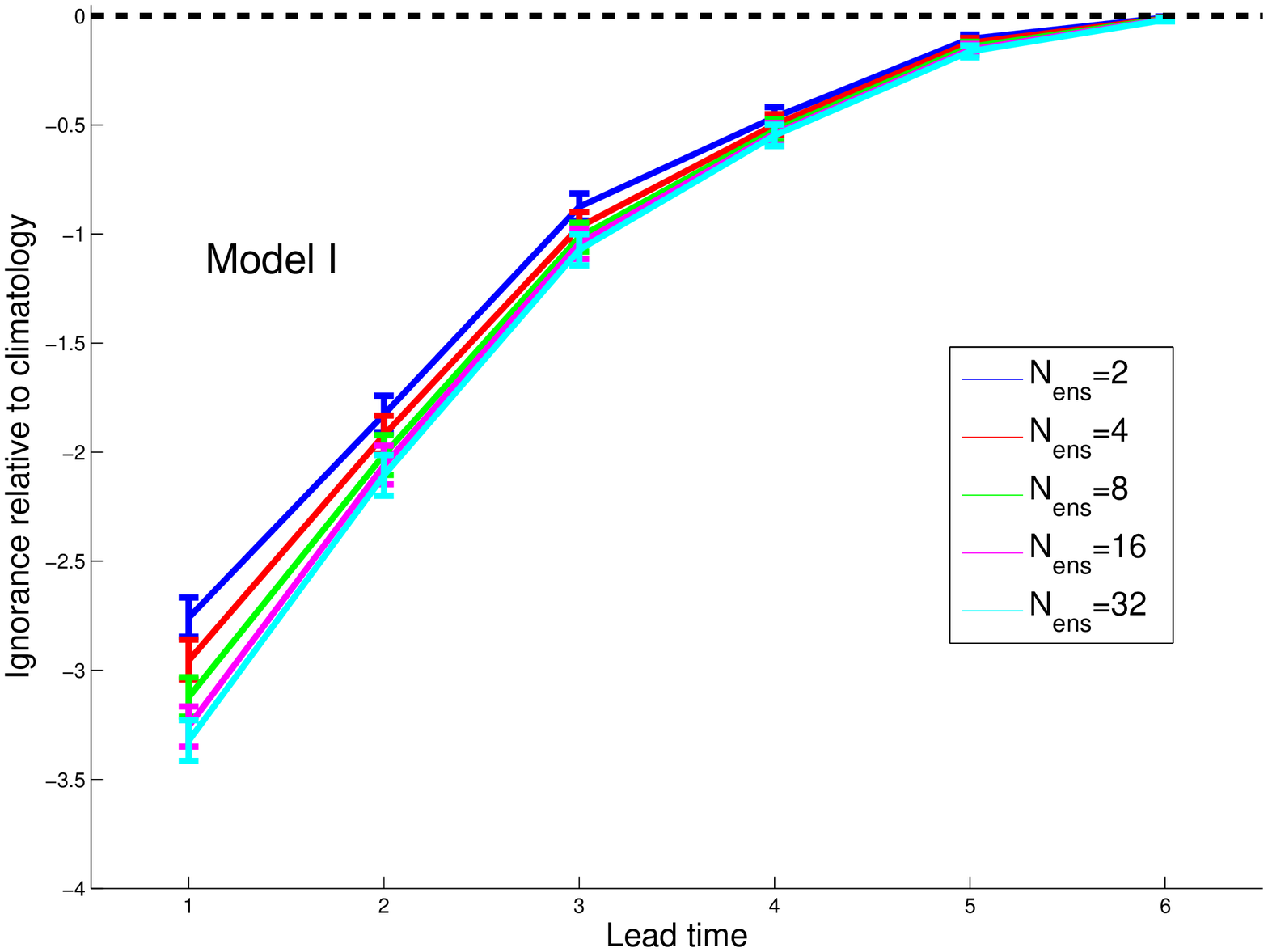, width=0.48\columnwidth, height=6cm}
  \epsfig{file=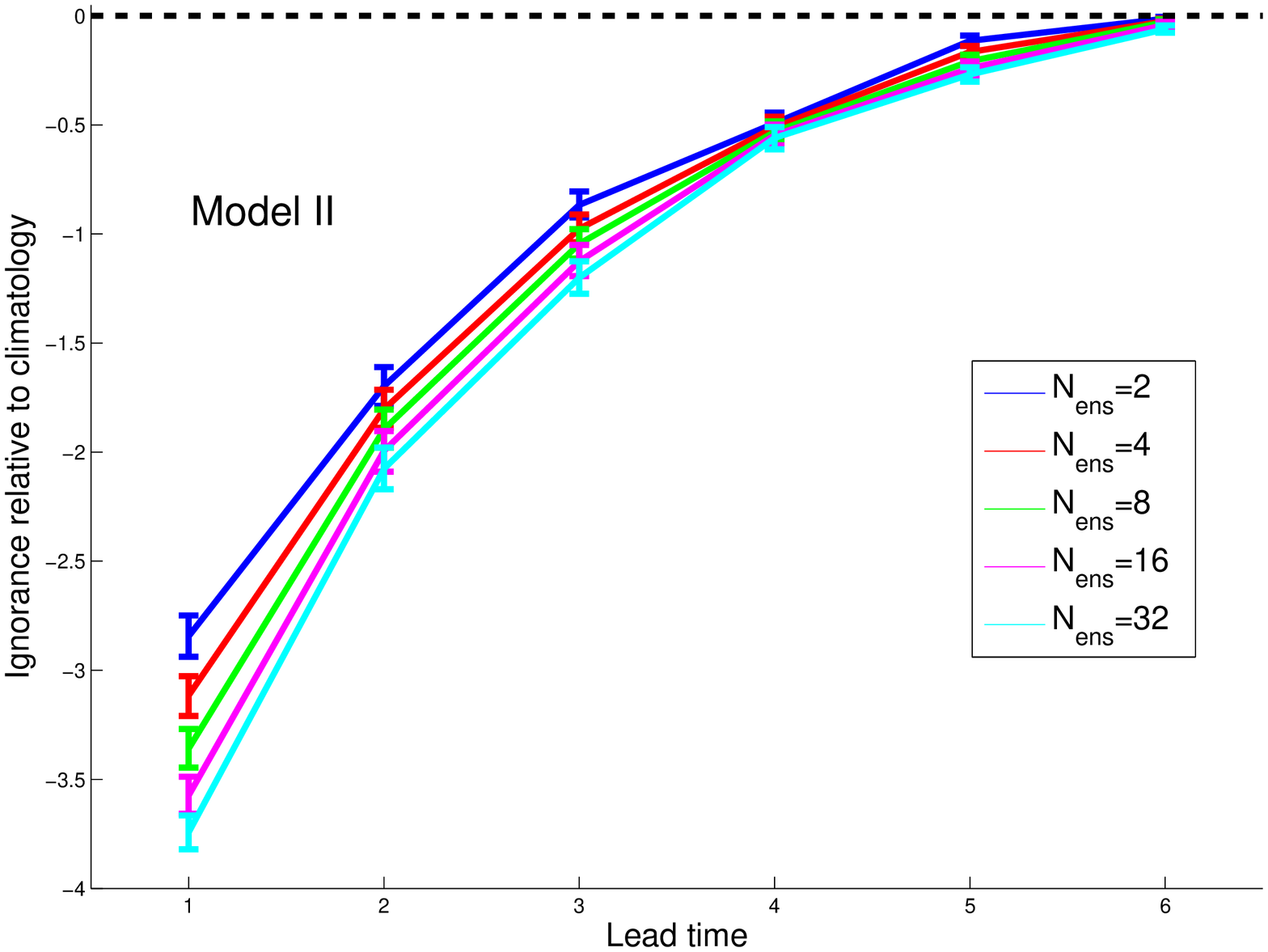, width=0.48\columnwidth, height=6cm}
}
\hbox{
  \epsfig{file=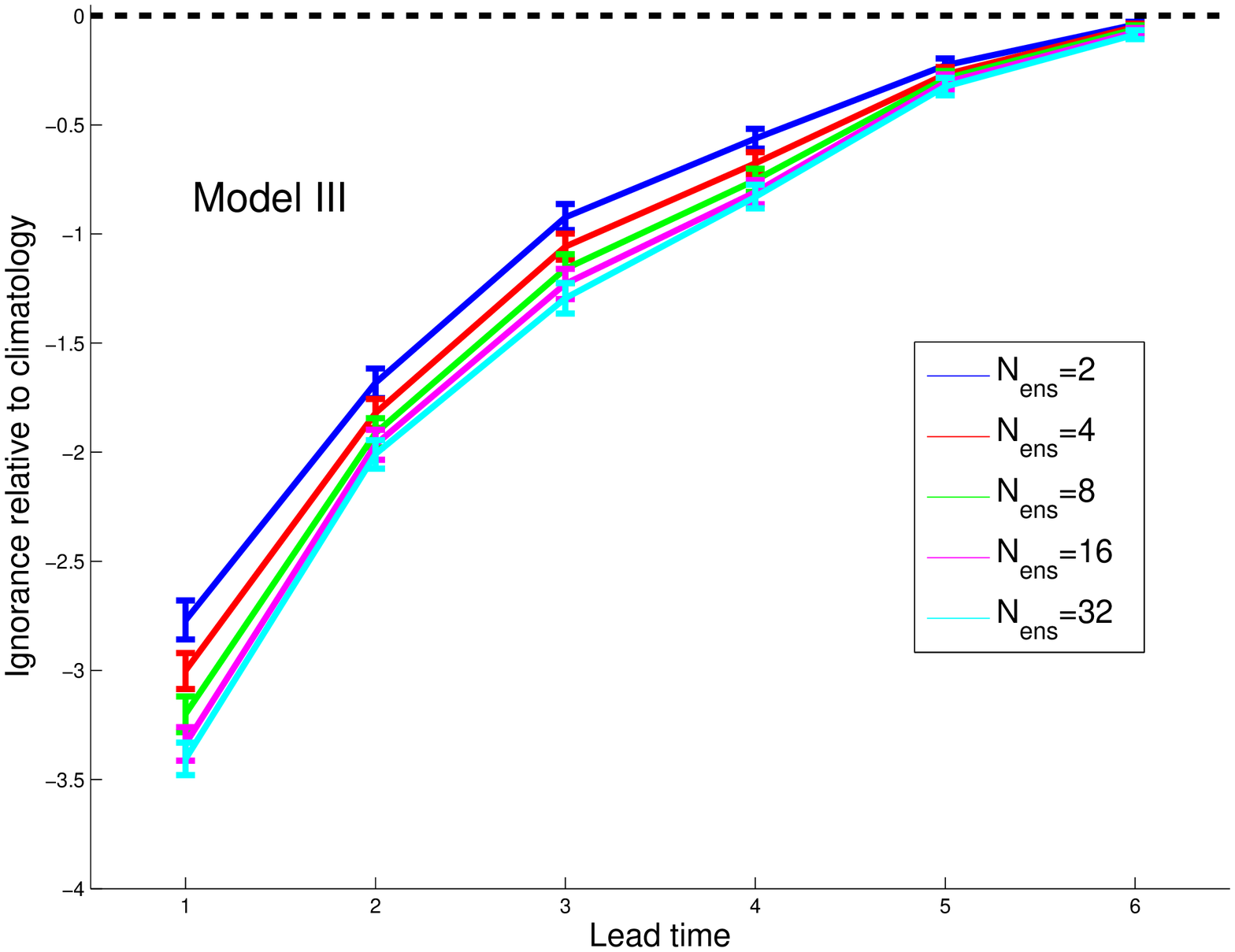, width=0.48\columnwidth, height=6cm}
  \epsfig{file=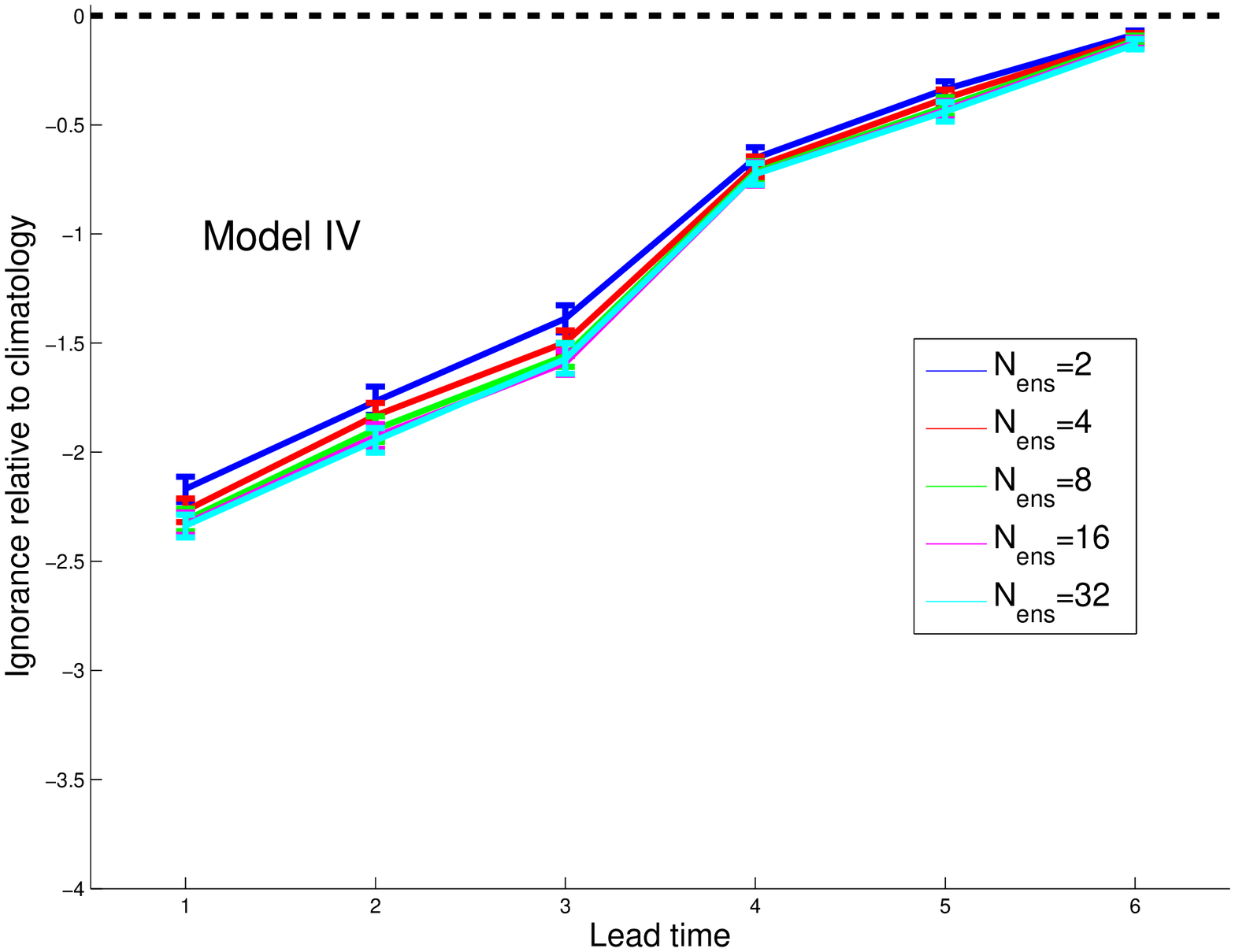, width=0.48\columnwidth, height=6cm}
}
\caption{The Ignorance score varies as the ensemble size increases for each model.}
\label{fig:ens_size}
\end{figure}

As the number of ensemble members increases, the true limits of the model structure become more apparent. Figure~\ref{fig:ens_size} shows forecast Ignorance varies as ensemble size increases. {Improvement from additional ensemble members can be noted, especially at shorter lead times.}

%
%
%

\section{Forecast System Design and Model Weighting when data are precious}


\subsection{Forecast with a large forecast-outcome archive}



\begin{figure}[!h]
\hbox{
  \epsfig{file=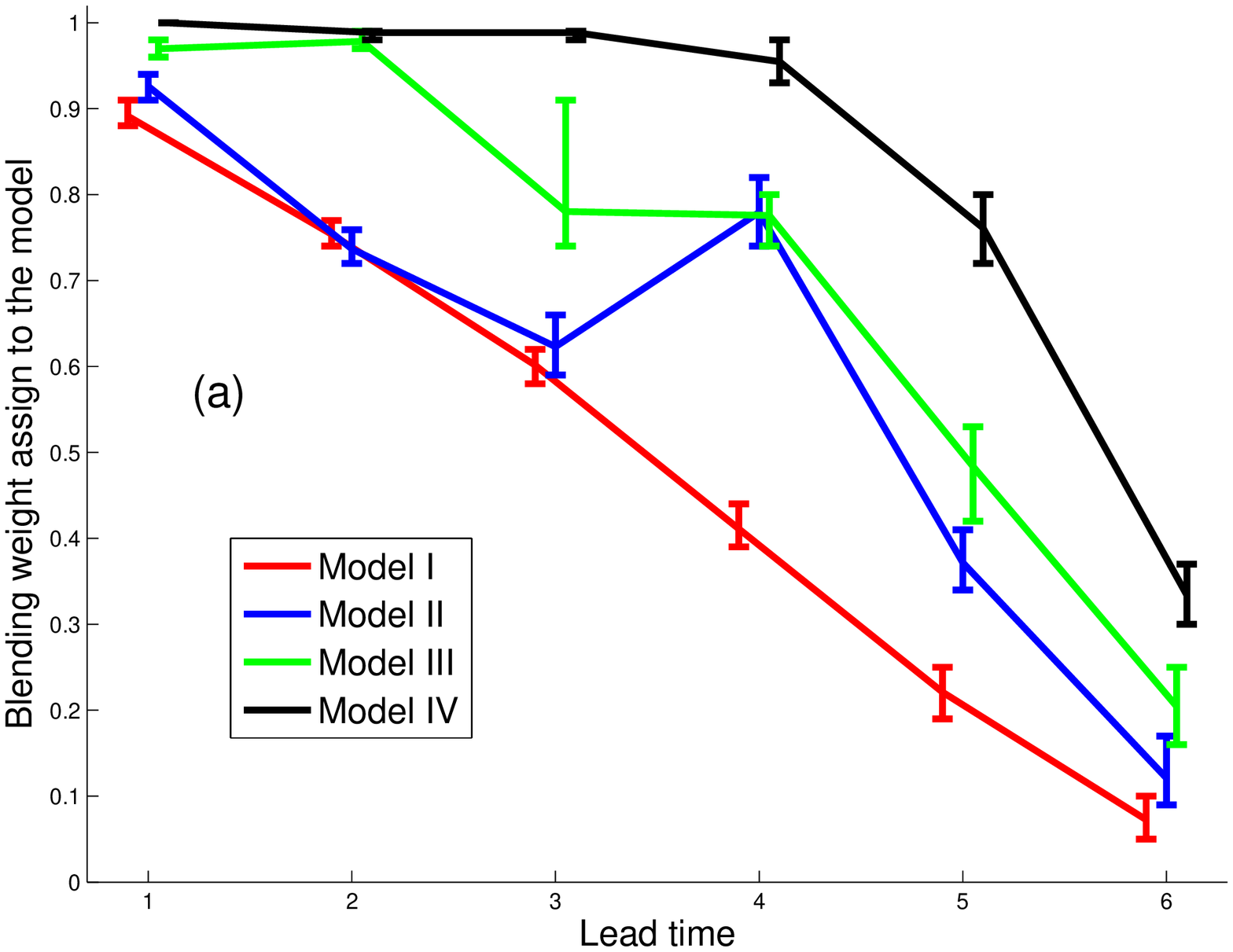, width=0.48\columnwidth, height=6cm}
  \epsfig{file=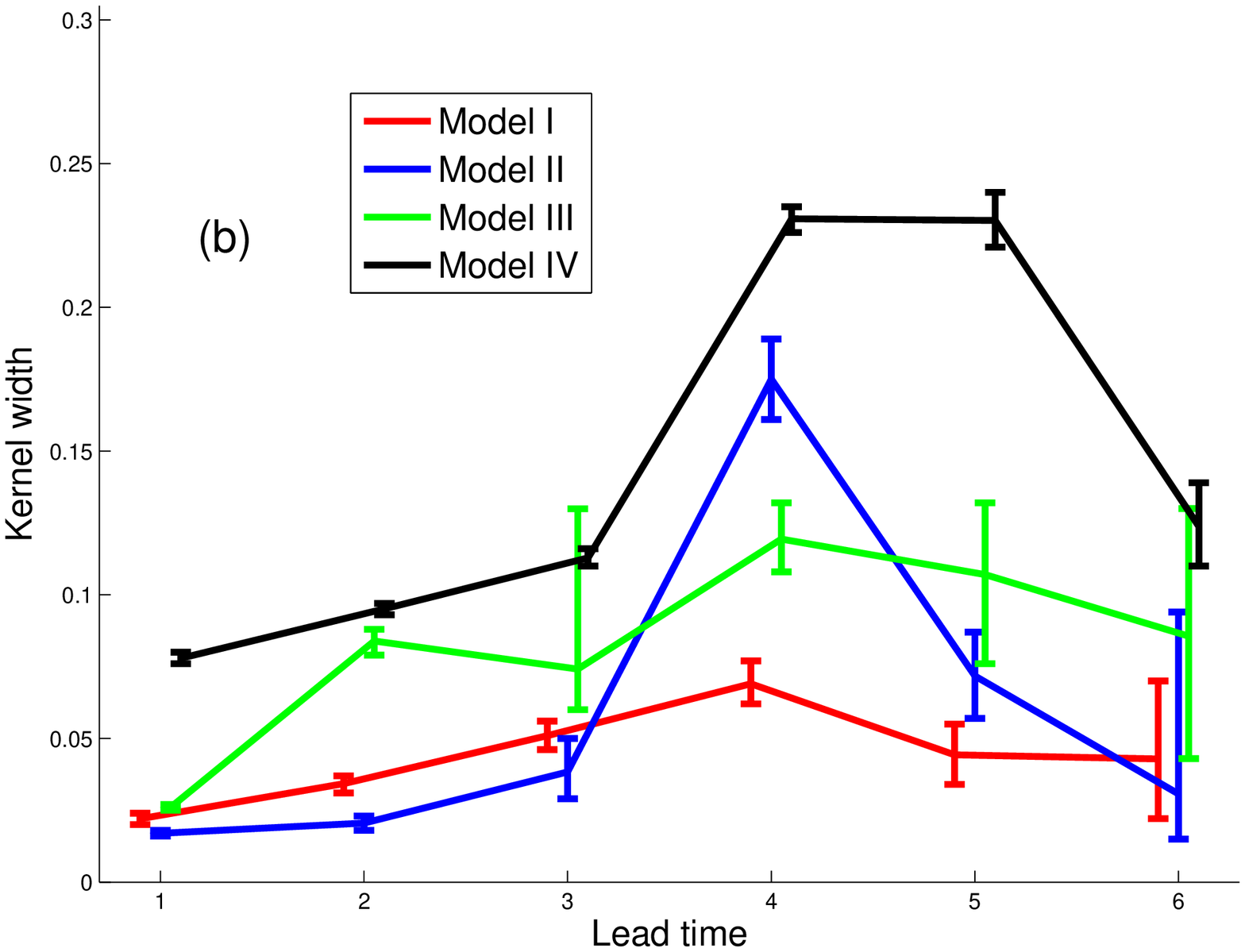, width=0.48\columnwidth, height=6cm}
}
\hbox{
  \epsfig{file=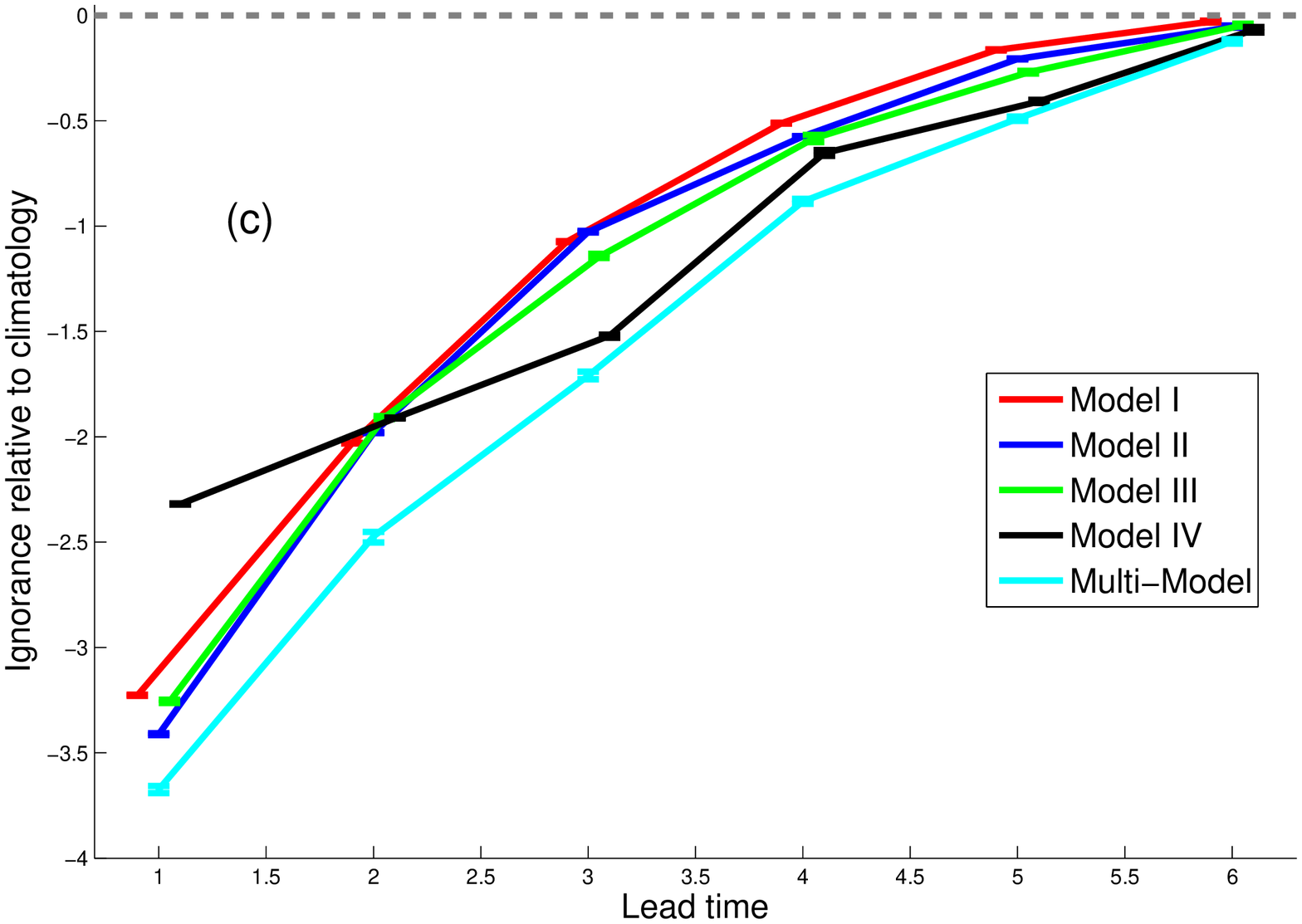, width=0.48\columnwidth, height=6cm}
  \epsfig{file=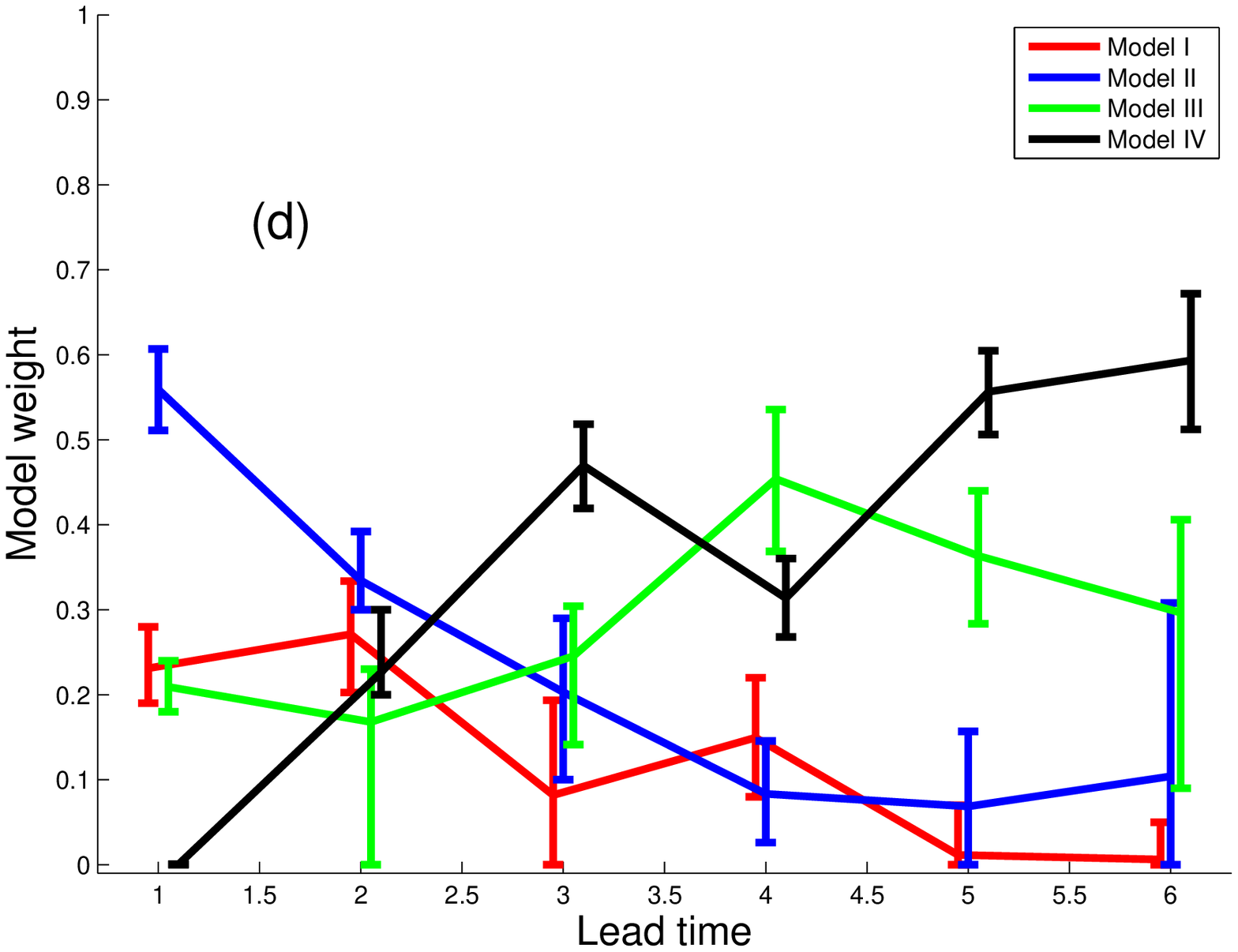, width=0.48\columnwidth, height=6cm}
}
\caption{Forecast Ignorance, climatology-blend weight assigned to the model, kernel width and weights assigned to each individual model are plotted as a function of lead time.}
\label{fig:largeset}
\end{figure}

As the size of the forecast-outcome archive, $N_a$, increases one expects robust results since large training sets and large testing sets can be considered. To examine this, 512 different training sets are produced, each contains 2048 forecast-outcome pairs. And for each archive, the kernel width $\sigma$ and climatology-blend weight $\alpha$ for each model forecasts are fitted at different lead time. Figures~\ref{fig:largeset}a and~\ref{fig:largeset}b show the fitted values of dressing parameters and climatology-blend weights. The error bars reflect the central $90^{th}$ percentile over 512 samples. The variation of the weight assigned to the model appears small. The variation of the fitted kernel width is small at short lead time and large at long lead time. Especially at lead time $6$, the fitted value for model IV has relatively large variation. This, however, does not indicate the estimate is not robust but suggests the Ignorance score function in the parameter space is relatively flat near the minimum. To demonstrate this the empirical Ignorance is calculated for each archive of kernel width and climatology-blend weight based on the same testing set which contains another 2048 forecast-outcome pairs. Figure~\ref{fig:largeset}c plots the Ignorance score and its $90^{th}$ percentile as a function of lead time. Notice the $90^{th}$ percentile ranges are very narrow all the time. 

There are many ways in which forecast distributions, generated from ensembles of individual model runs can be combined to produce a single probabilistic multi-model forecast distribution. One approach may be to assign equal weight to each model and simply sum the distributions generated from each model to obtain a single probabilistic distribution (see \cite{Hagedorn}). In general, different forecast models do not provide equal amounts of information, one may want to weight the models according to some measure of past performance, see for example~\cite{Rajagopalan,Doblas-Reyes}. The combined multi-model forecast is the weighted linear sum of the constituent distributions,
\begin{eqnarray}
 \label{eq:weight}
    p_{mm}=\sum_{i}\omega_{i}p_{i},
\end{eqnarray}
\noindent where the $p_{i}$ is the forecast distribution from model $i$ and $\omega_{i}$ its weight, with $\sum_{i}\omega_{i}=1$. The weighting parameters may be chosen by minimizing the Ignorance score for example, although fitting $\omega_{i}$ in this way can be costly and is typically complicated by different models sharing information. {And, of course, the weights of individual models are expected to vary as a function of lead time.} 


To avoid ill fitting model weights, a simple iterative method to combine models is used below instead of fitting all the weights simultaneously. For each lead time, the best (in terms of Ignorance) model is first combined with the second best model to form a combined forecast distribution (by assigning weights to both models). The combined forecast distribution is then combined with the third best model to update the combined forecast distribution. Repeat this process until the worst model is considered. Figure~\ref{fig:largeset}d shows the weights assigned to each model as a function of lead time. The cyan line in Figure~\ref{fig:largeset}c shows the variation of Ignorance score for the multi-model forecast given those estimated model weights is very small.

\subsection{Forecast with a small forecast-outcome archive}

\begin{figure}[!h]
\hbox{
  \epsfig{file=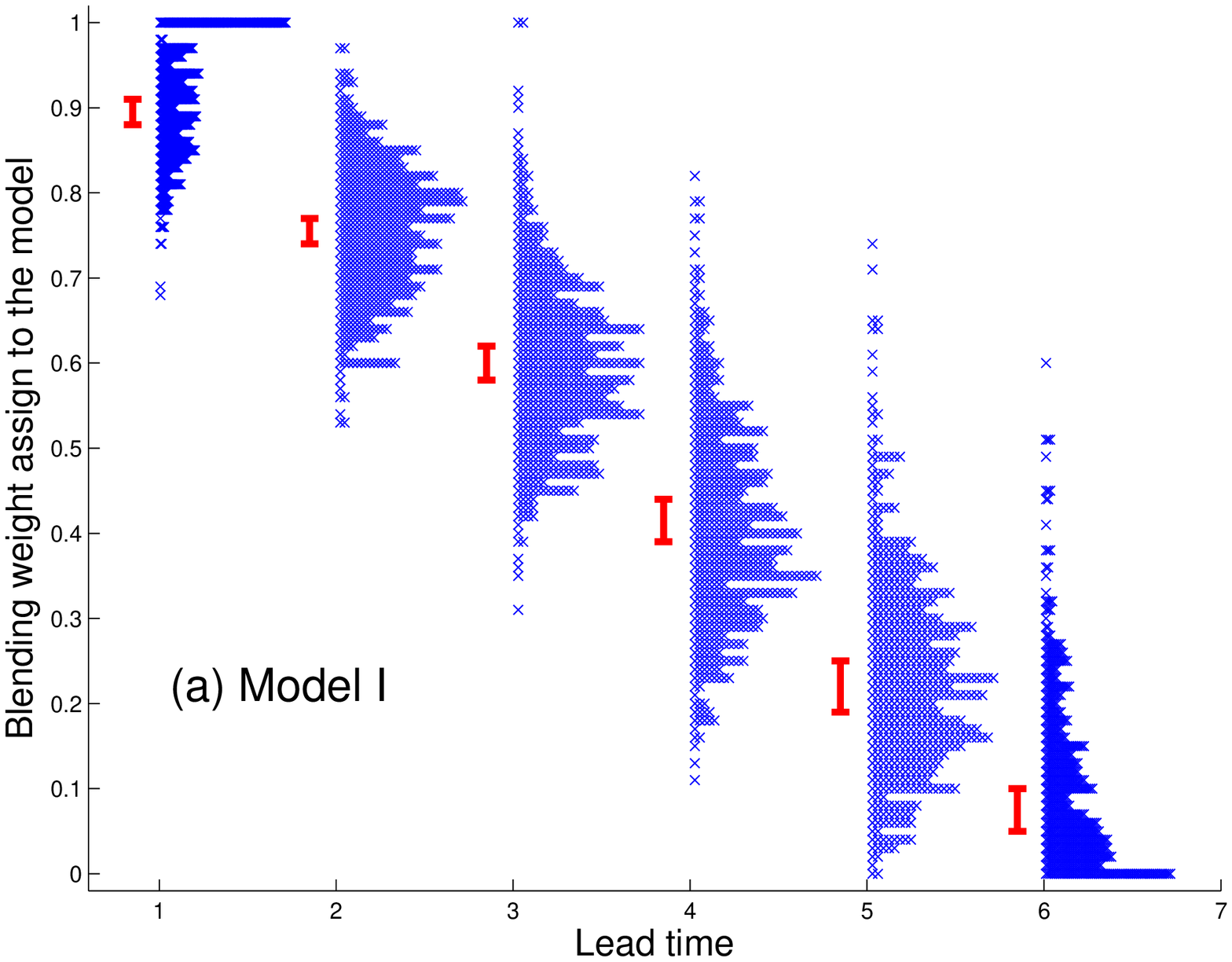, width=0.48\columnwidth, height=6cm}
  \epsfig{file=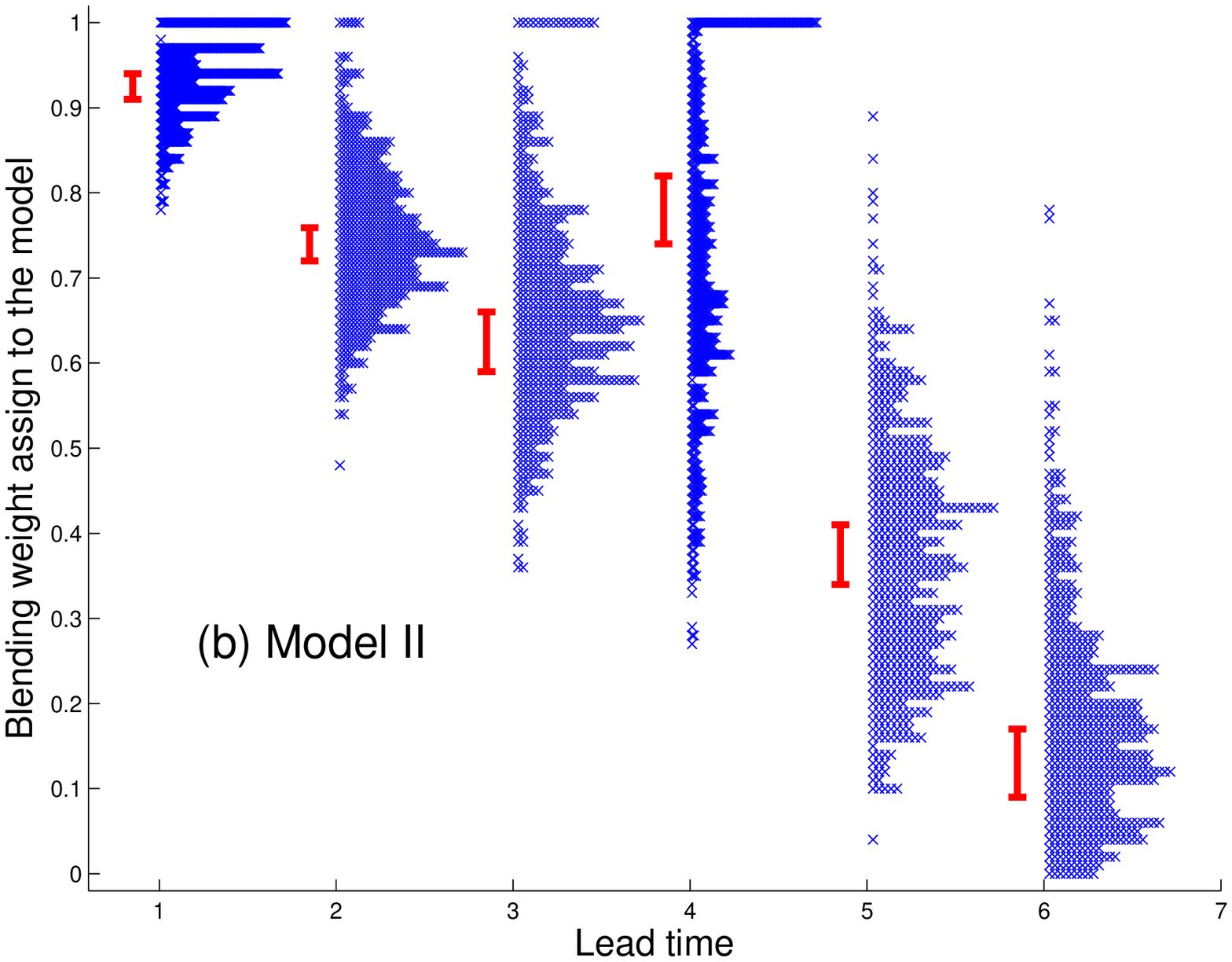, width=0.48\columnwidth, height=6cm}
}
\hbox{
  \epsfig{file=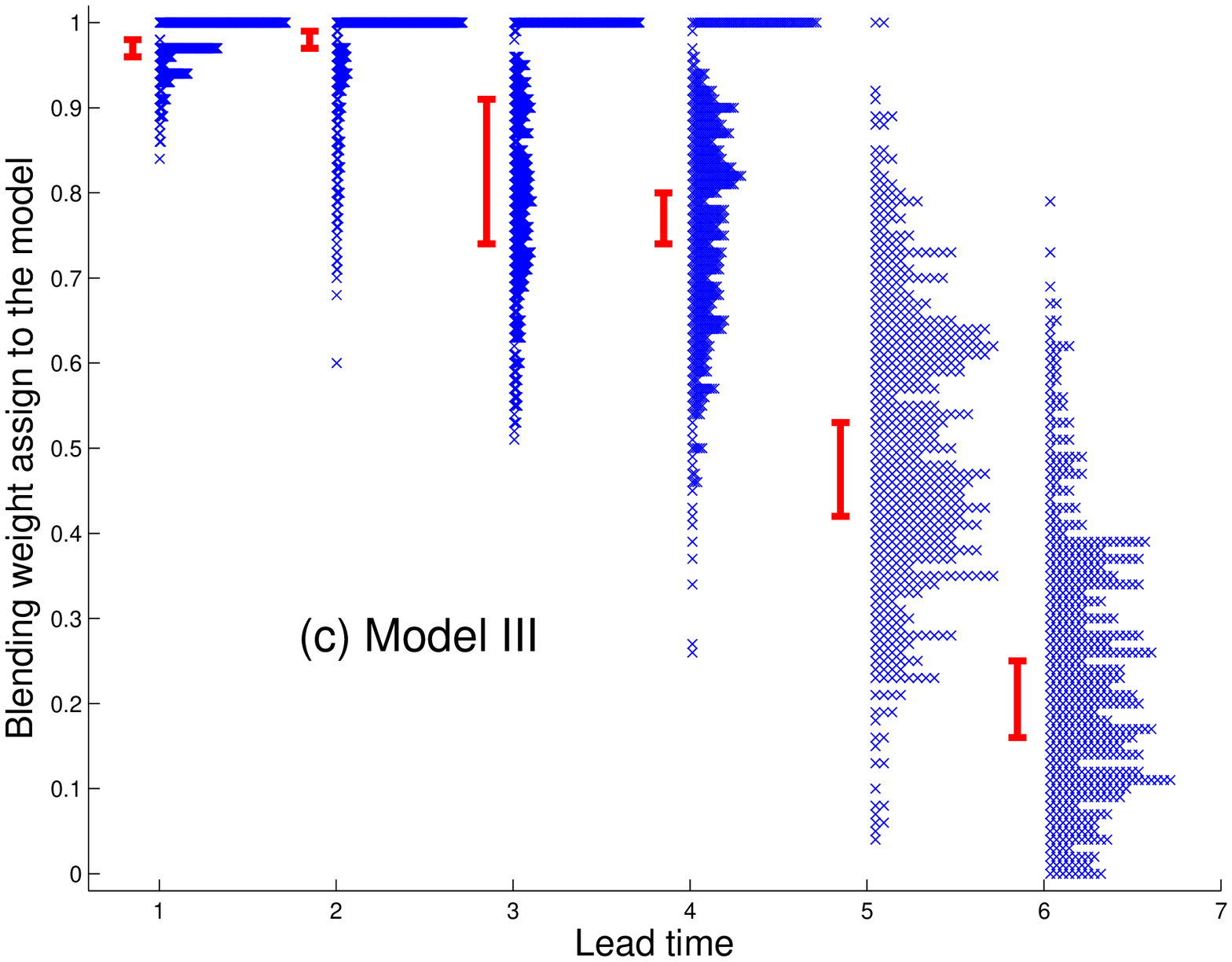, width=0.48\columnwidth, height=6cm}
  \epsfig{file=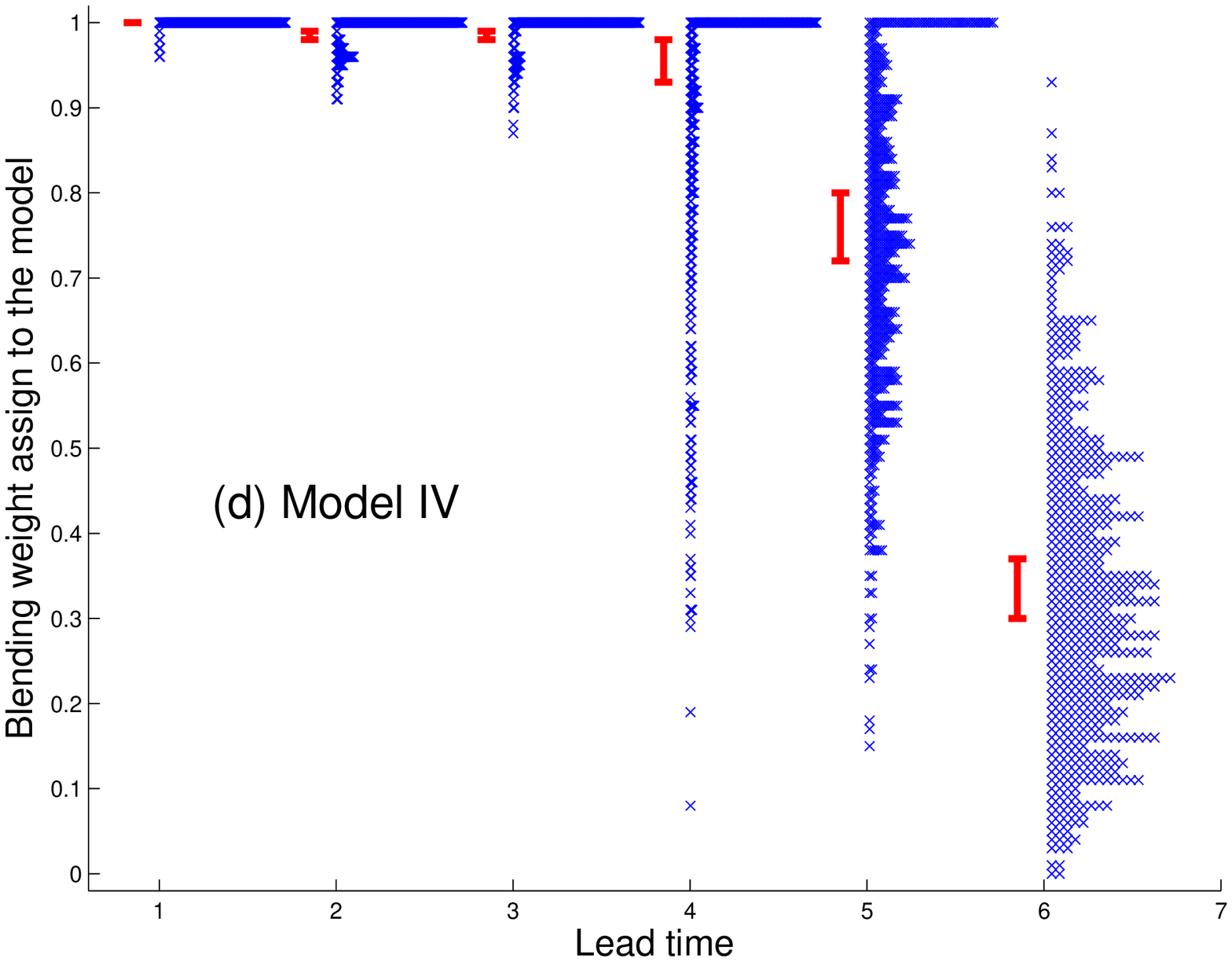, width=0.48\columnwidth, height=6cm}
}
\caption{Climatology-blend weights assigned to each model. The red bars are the $95^{th}$ percentile range of the fitted weights based on 512 forecast-outcome archives, each contains 2048 forecast-outcome pairs. The blue crosses represent the histogram of the fitted weights based on 512 forecast-outcome archives, each contains only 40 forecast-outcome pairs.}
\label{fig:alpha_small}
\end{figure}
\begin{figure}[!h]
\hbox{
  \epsfig{file=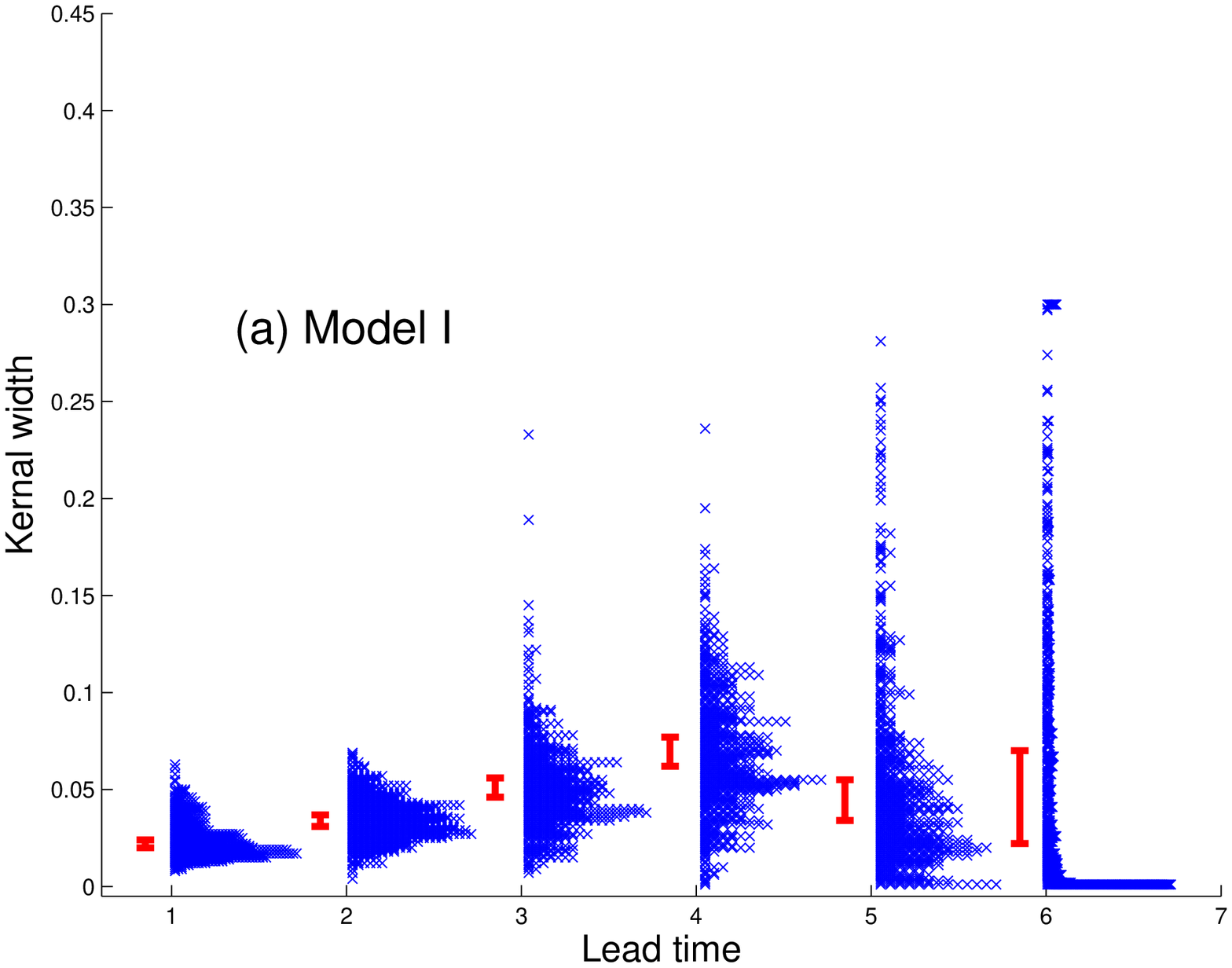, width=0.48\columnwidth, height=6cm}
  \epsfig{file=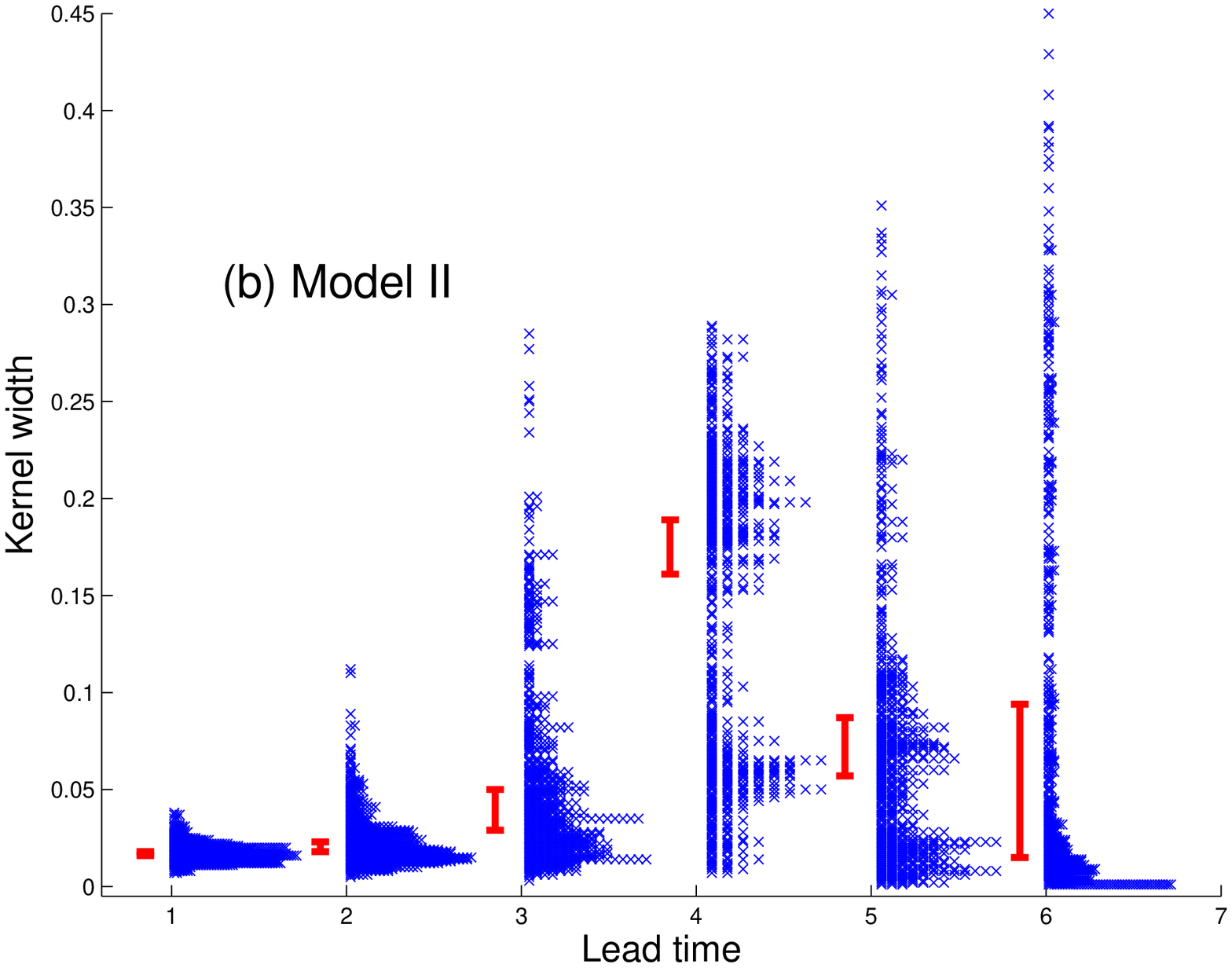, width=0.48\columnwidth, height=6cm}
}
\hbox{
  \epsfig{file=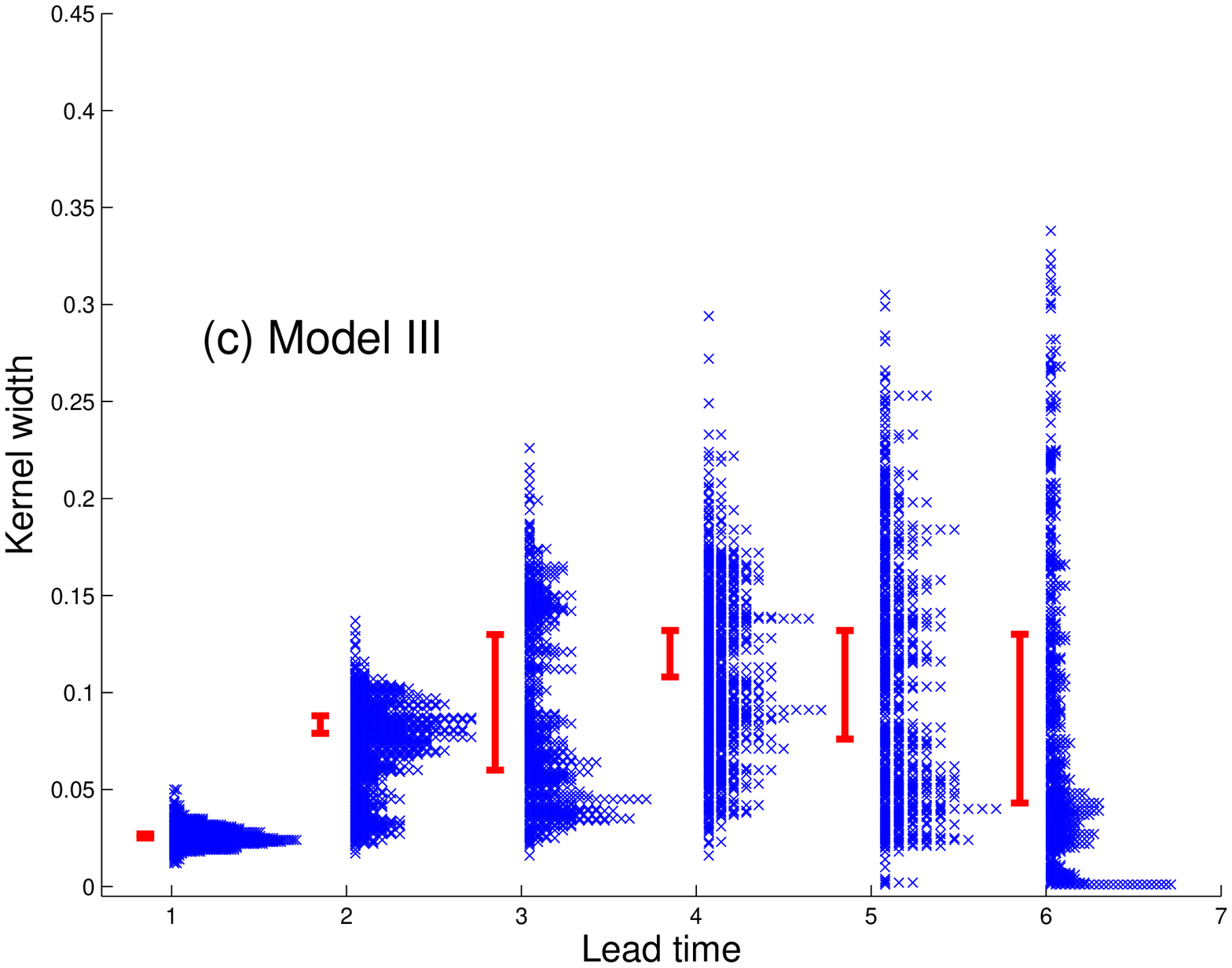, width=0.48\columnwidth, height=6cm}
  \epsfig{file=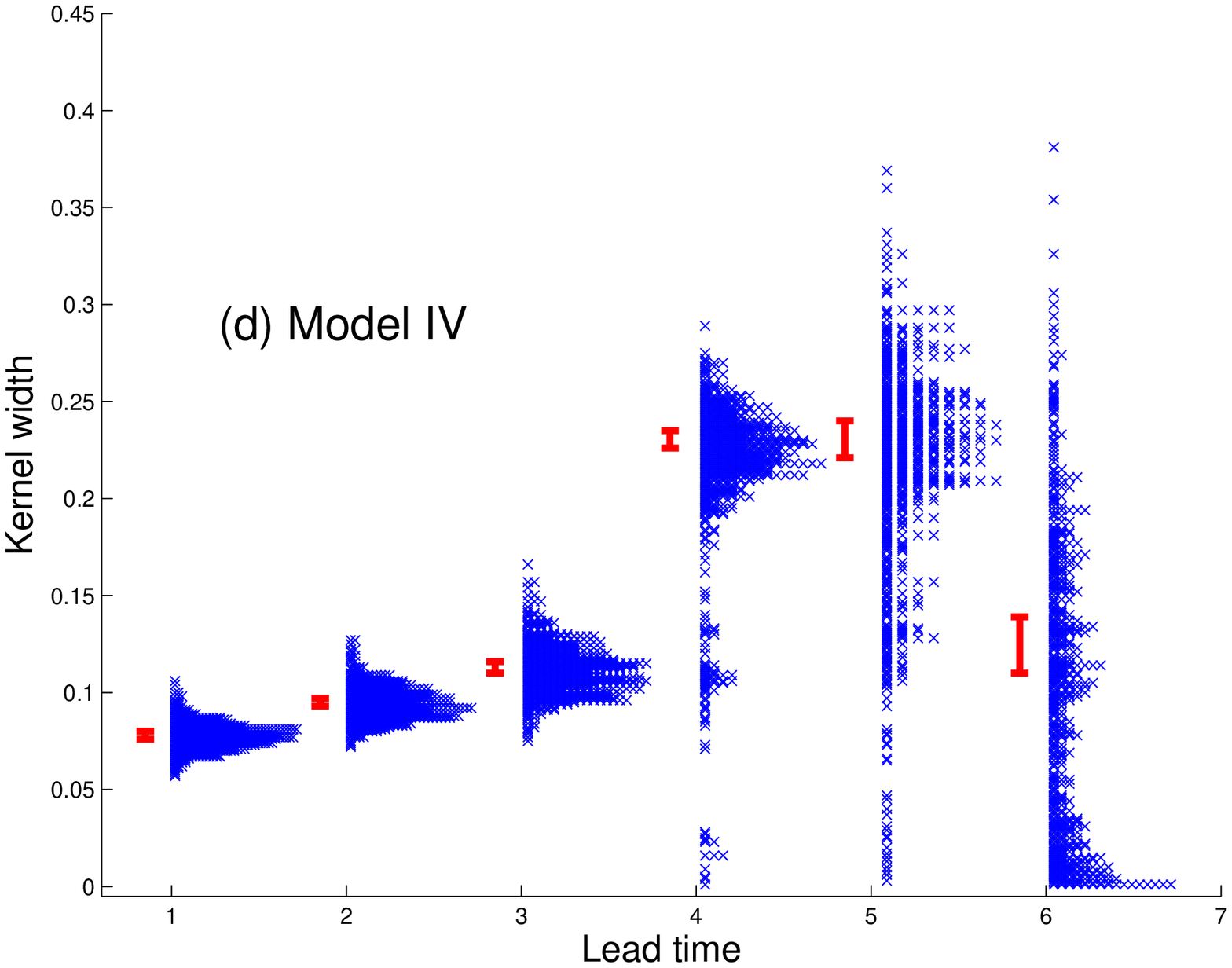, width=0.48\columnwidth, height=6cm}
}
\caption{Kernel width of each model forecasts. The red bars are the $95^{th}$ percentile range of the fitted kernel width based on 512 forecast-outcome archives, each contains 2048 forecast-outcome pairs. The blue crosses represent the histogram of the fitted kernel width based on 512 forecast-outcome archives, each contains only 40 forecast-outcome pairs.}
\label{fig:segma_small}
\end{figure}

When given a small forecast-outcome archive (e.g. $\sim 40$ year seasonal forecast-outcome archive), one doesn't have the luxury of exploring a large collection of independent training and testing sets. Cross validation is often approached by adopting a leave-one-out approach. The robustness of fitting in such cases is of more concern. To examine such robustness, a large number of forecast-outcome archives are considered, each archive contains the same numbers of forecast-outcome pairs. For each archive, the parameter values are fitted via leave-one-out cross-validation. The distribution of fitted values over these small forecast-outcome archives are then compared with the fitted value from $N_a=2048$ large forecast-outcome archives above. Figure~\ref{fig:alpha_small} plots the histograms of the fitted climatology-blend weights given 512 forecast-outcome archives each contain $N_a=40$ forecast-outcome pairs. Notice that in most of the cases the distributions are very wide although they cover the value fitted given the large training set. There are some cases in which about 90 percent of the estimates are larger or smaller than the values fitted by large archive, e.g. lead time 1 of Model I and Model II, lead time 4 of Model III and lead time 5 of Model IV. It appears that {\it the robustness of fitting varies with lead time and the model}. For shorter lead time however the weights are more likely over fitted and for longer lead time the weights are more likely under fitted. This is because at short lead time the model forecasts are relatively good; only a few forecasts are worse than the climatological forecast, and small forecast-outcome archives may not contain any model busts and so over estimate the weights. The longer lead time case can be similarly explained. Figure~\ref{fig:segma_small} plots the histogram of fitted kernel widths. Again observe there is much larger variation of the estimates than fitting with large forecast-outcome archives.

\begin{figure}[!h]
\hbox{
  \epsfig{file=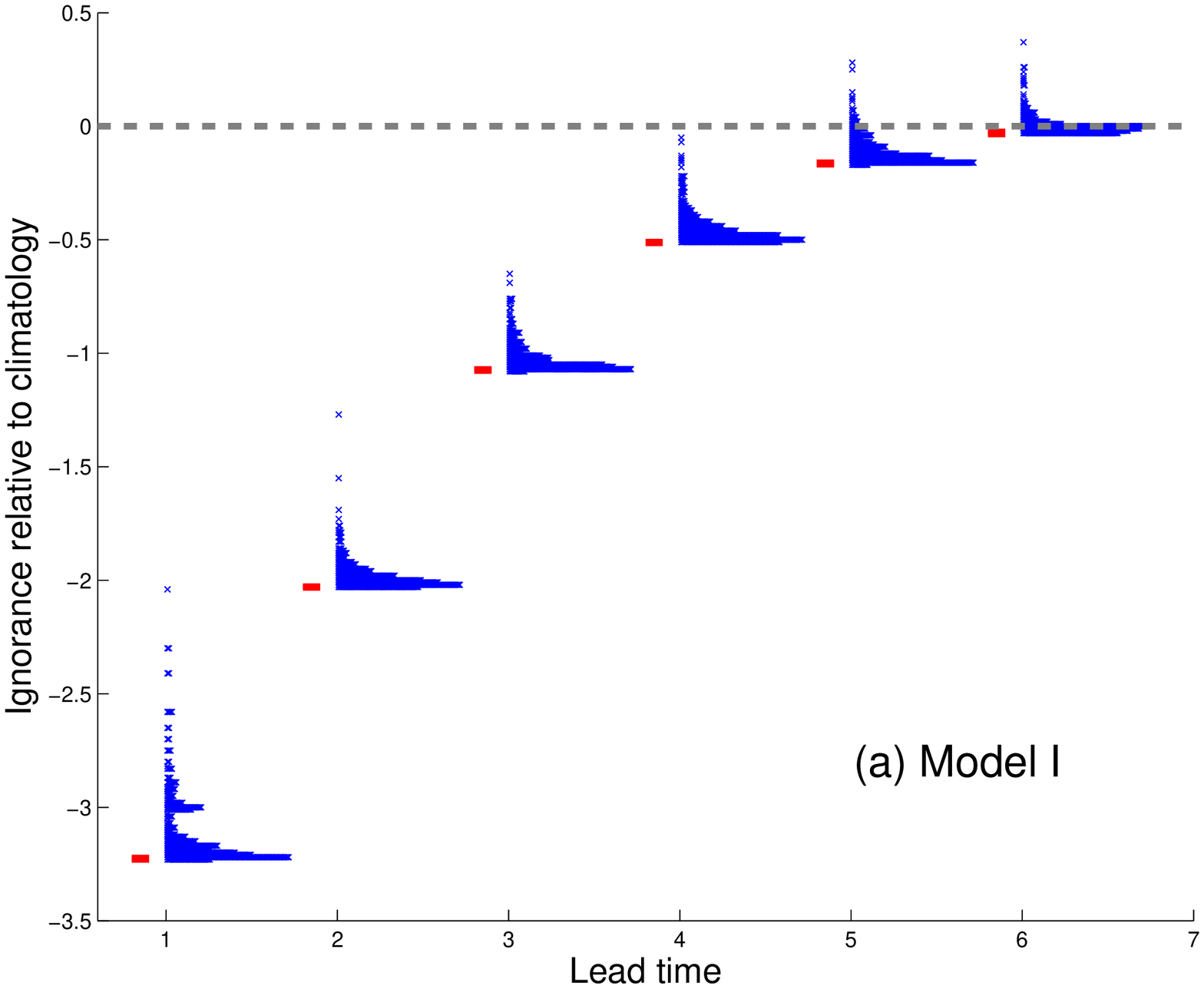, width=0.48\columnwidth, height=6cm}
  \epsfig{file=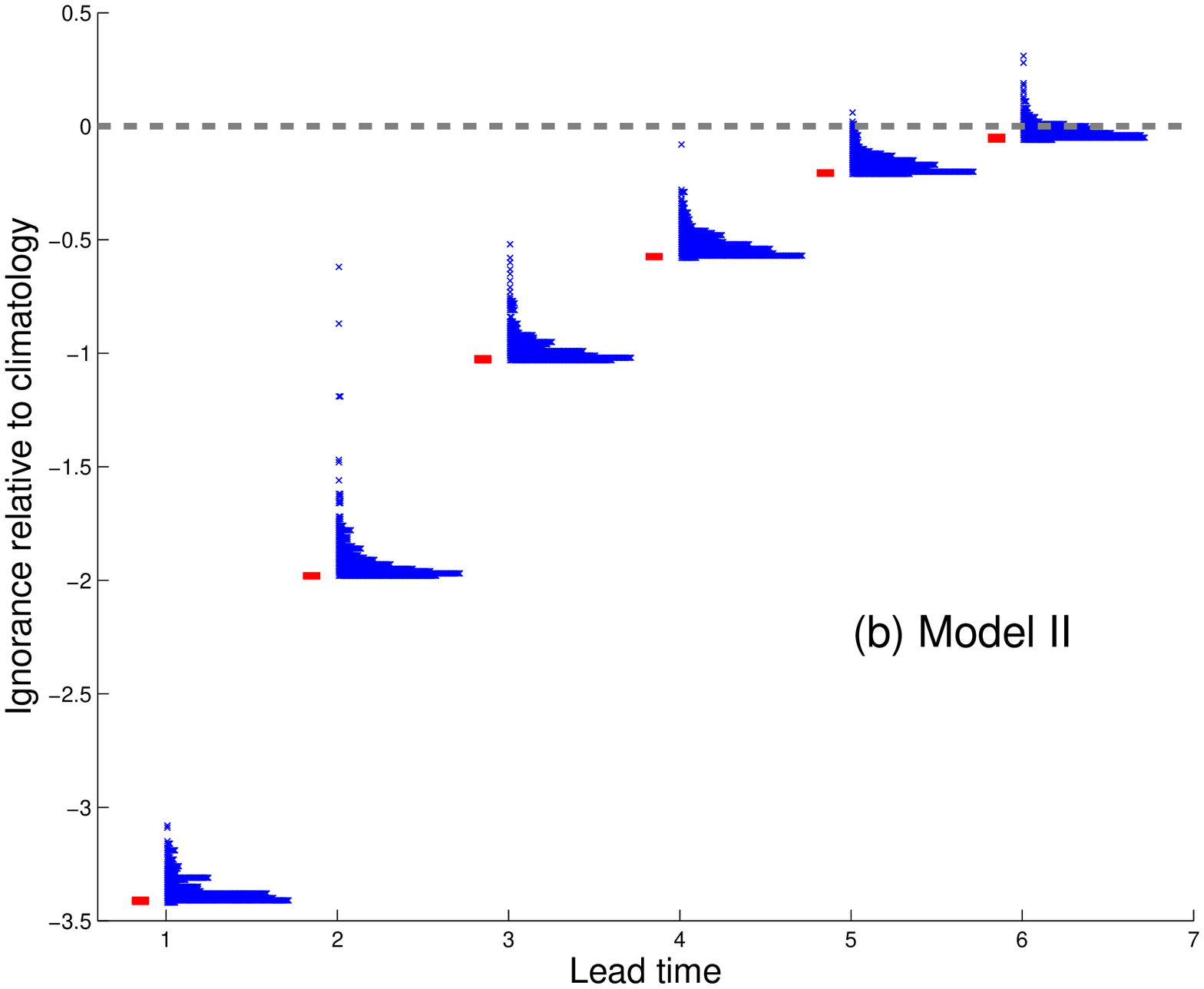, width=0.48\columnwidth, height=6cm}
}
\hbox{
  \epsfig{file=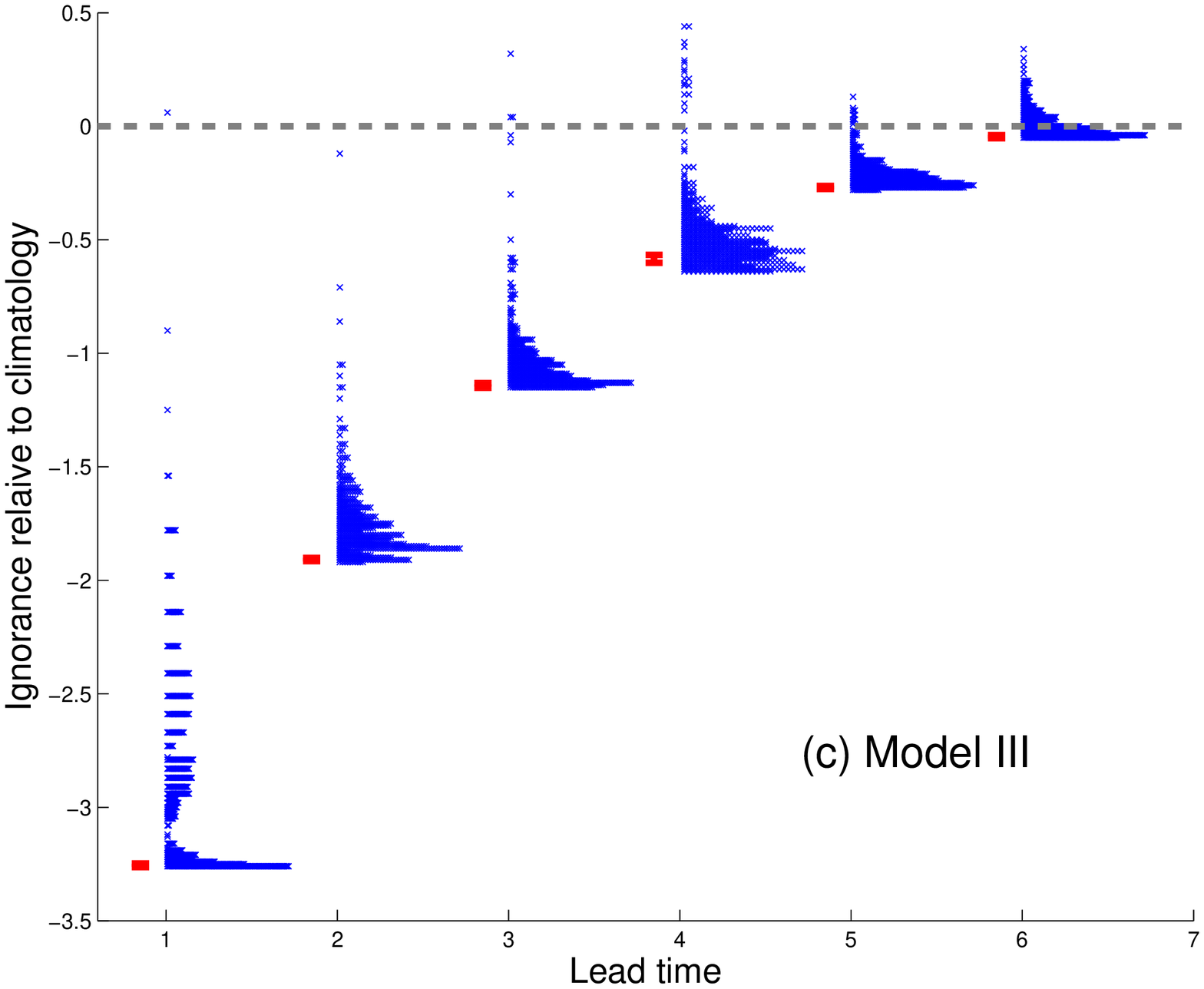, width=0.48\columnwidth, height=6cm}
  \epsfig{file=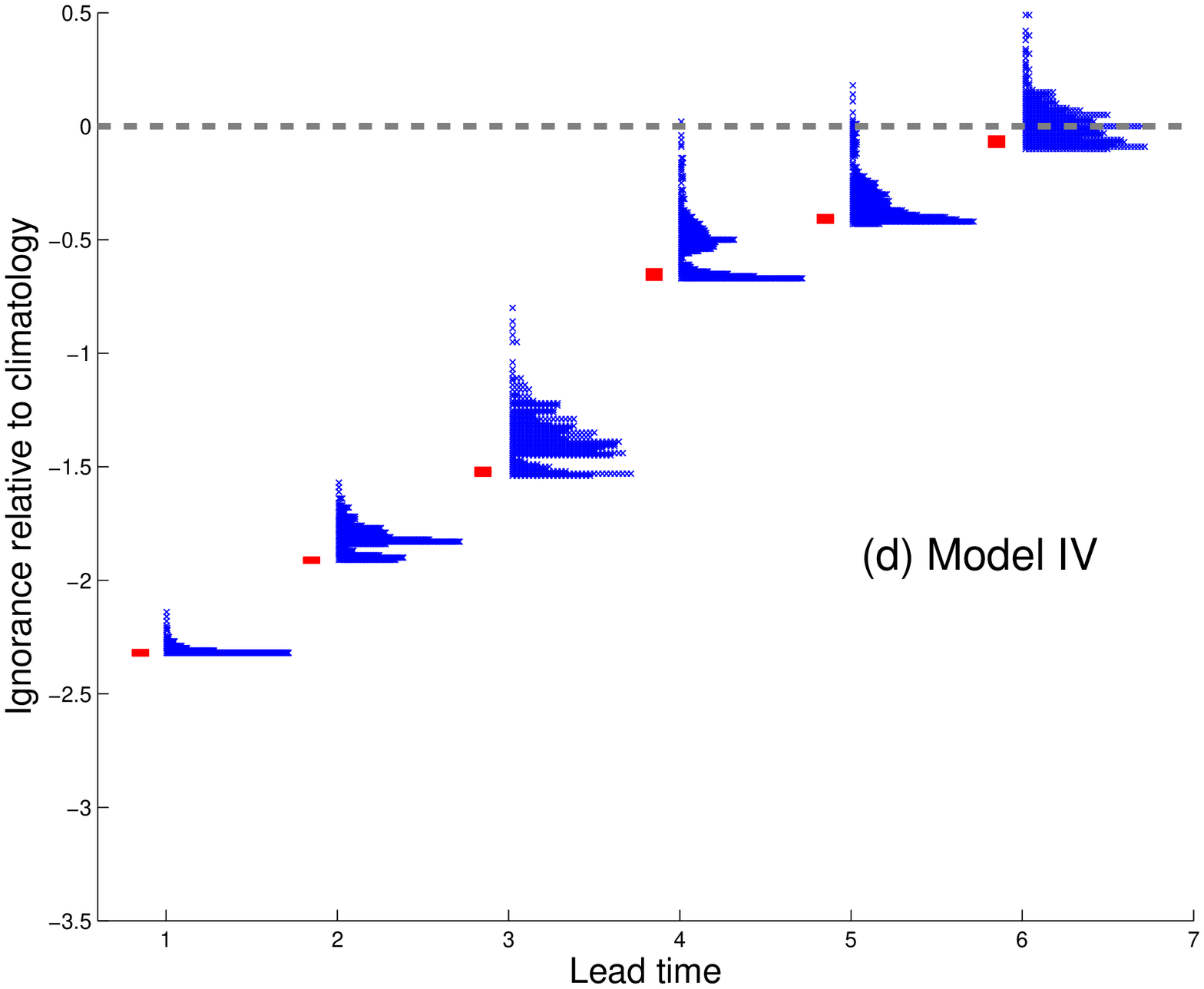, width=0.48\columnwidth, height=6cm}
}
\caption{Ignorance score of each model forecasts. The red bars are the $95^{th}$ percentile range of Ignorance score calculated based on a testing set containing 2048 forecast-outcome pairs, using the climatology-blend weights and kernel widths fitted based on 512 forecast-outcome archives, each contains 2048 forecast-outcome pairs. The blue crosses represent the histogram of Ignorance score calculated based on the same testing set but using the climatology-blend weights and kernel widths based on 512 forecast-outcome archives, each contains only 40 forecast-outcome pairs.}
\label{fig:Ign_small}
\end{figure}

\begin{figure}[!h]
\hbox{
  \epsfig{file=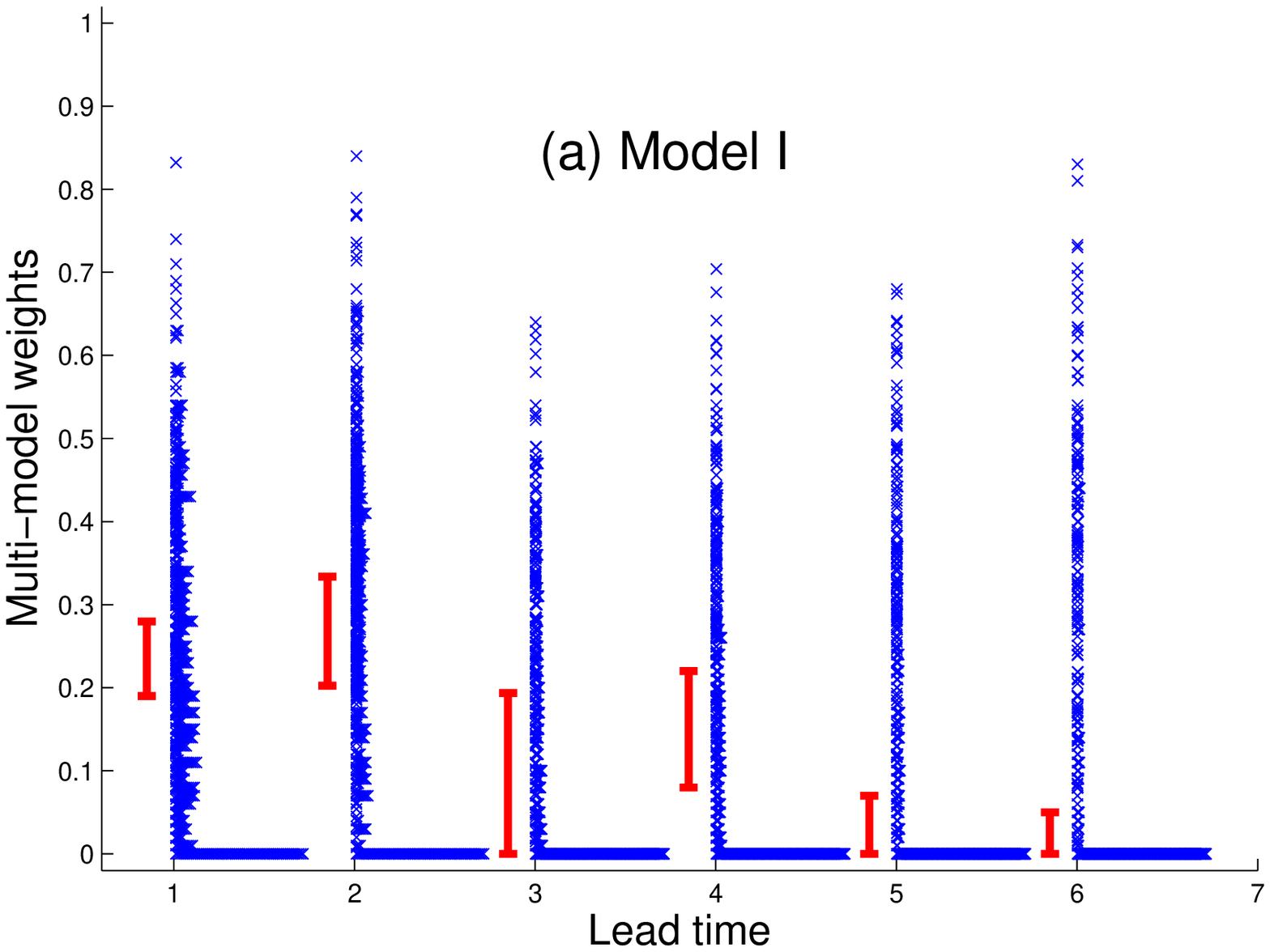, width=0.48\columnwidth, height=6cm}
  \epsfig{file=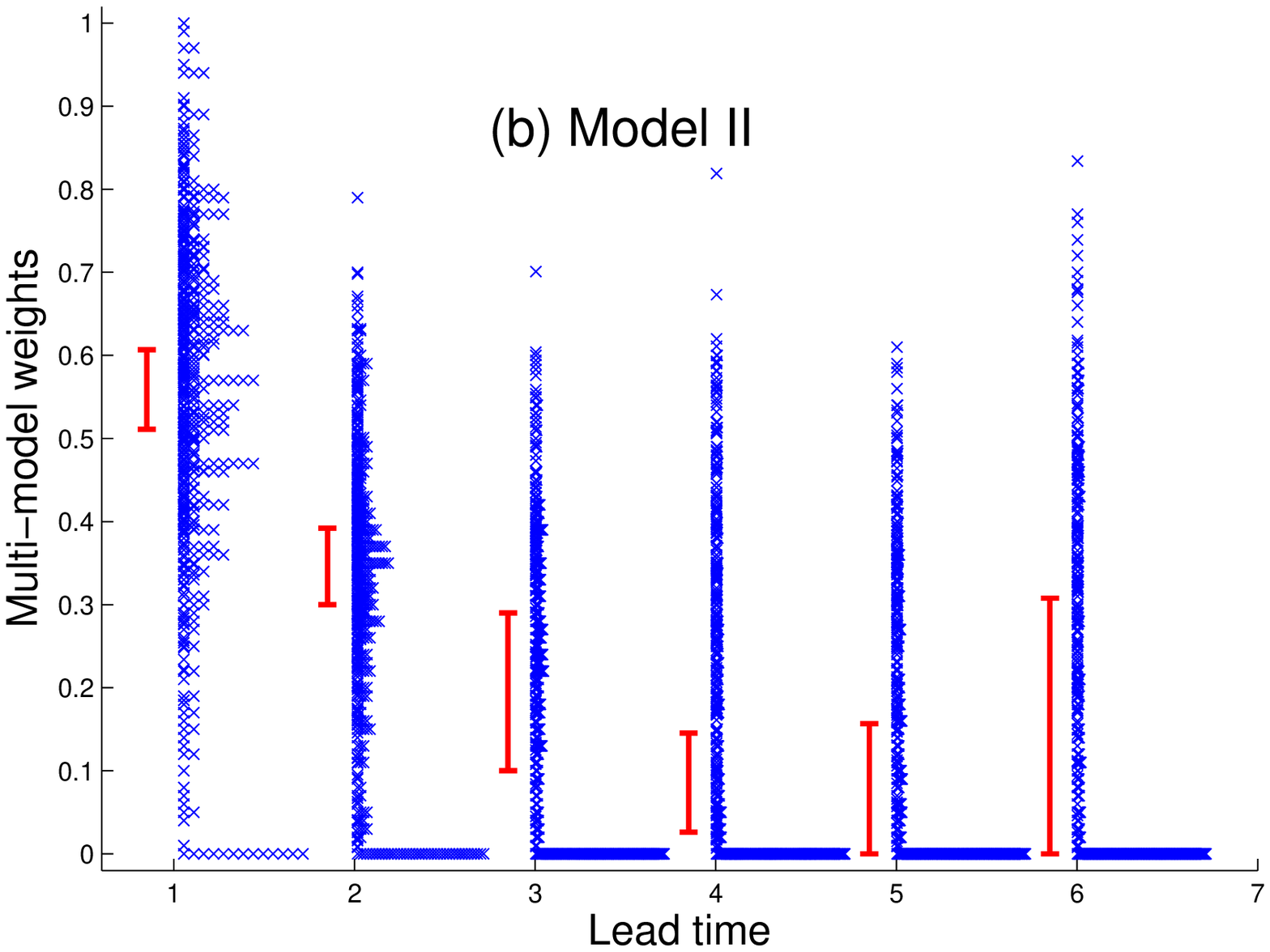, width=0.48\columnwidth, height=6cm}
}
\hbox{
  \epsfig{file=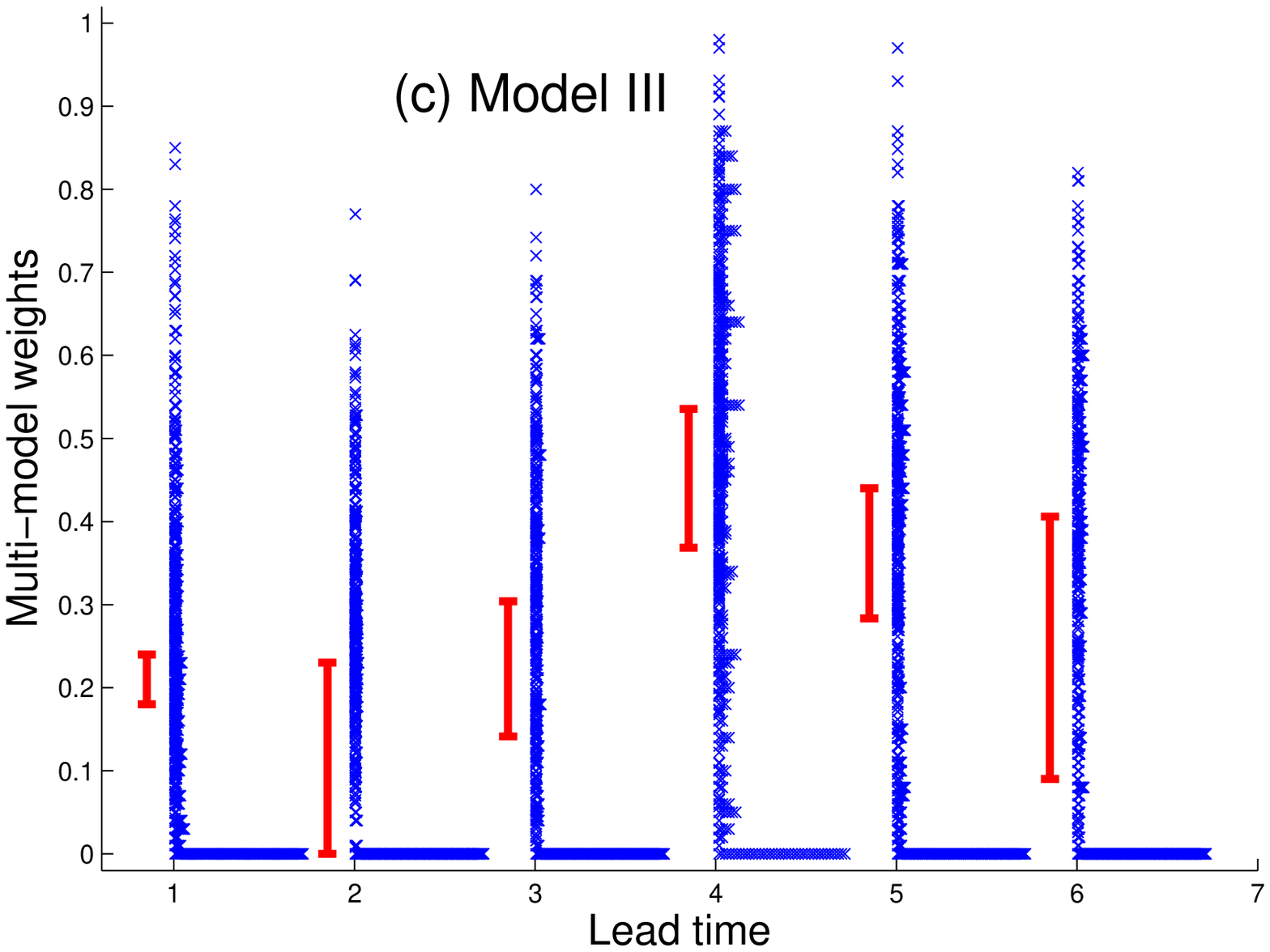, width=0.48\columnwidth, height=6cm}
  \epsfig{file=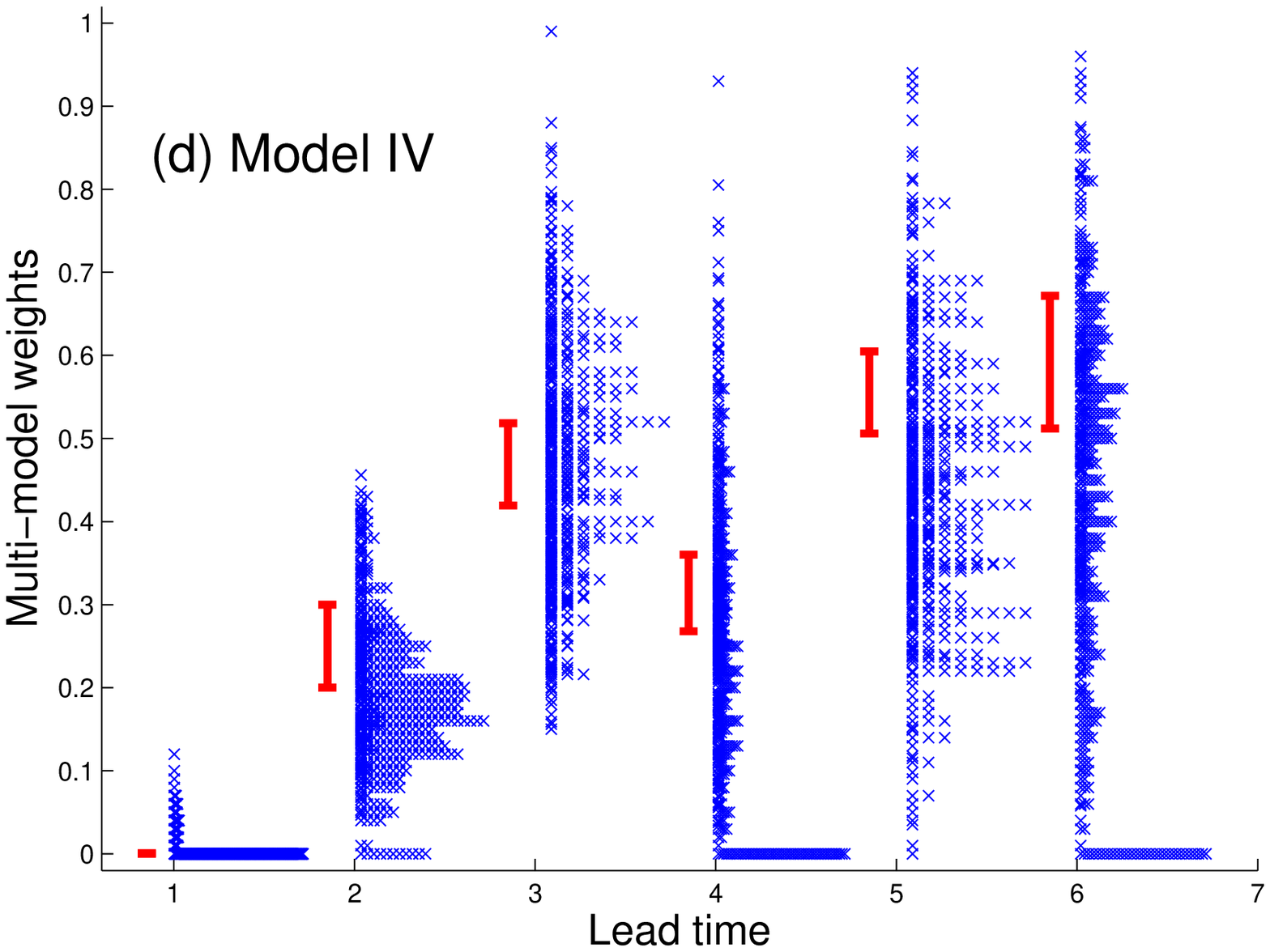, width=0.48\columnwidth, height=6cm}
}
\caption{Multi-model weights of each model forecasts. The red bars are the $95^{th}$ percentile range of model weights calculated based on a testing set containing 2048 forecast-outcome pairs, using the climatology-blend weights and kernel widths fitted based on 512 forecast-outcome archives, each contains 2048 forecast-outcome pairs. The blue crosses represent the histogram of model weights calculated based on the same testing set but using the climatology-blend weights and kernel widths based on 512 forecast-outcome archives, each contains only 40 forecast-outcome pairs.}
\label{fig:Weights_small}
\end{figure}

Poor estimation of the kernel width and climatology-blend weight will cause the forecast to lose skill. Given the 512 fitted kernel width and climatology-blend weights, the Ignorance scores for them is calculated over the same testing set of 2048 forecast-outcome pairs. Figure~\ref{fig:Ign_small} plots the histogram of the Ignorance score for each model. Using parameters fitted with small archives often results in significant degrading ($\sim 1$ bit) of the Ignorance score of the forecasts. Correctly blended with the climatological distribution would yield a forecast score which, in expectation, is never worse than the climatology; when the blending parameter is determined using the small archive, however, the relative Ignorance can be worse than climatology out of sample at long lead time (see for example in Figure~\ref{fig:Ign_small}). {Figure~\ref{fig:Weights_small} plots the histogram of multi-model weights. Clearly the variation of the model weights based on small archive are much larger, weights of zero are often assigned to models forecast which contain useful information. }

\section{Multi-model vs single best model}

It is sometimes said that a multi-model ensemble forecast is more skillful than any of its constituent single-model ensemble forecasts (see~\cite{Palmer04,Hagedorn,Bowler,Weigel,Weisheimer,Alessandri}). A common ``explanation" for this is the claim that the multi-model ensemble forecast reduces an apparent overconfidence in any one model (see for example~\cite{Weigel,Weisheimer,Alessandri}). As shown in section 6, single model SAP forecast systems are typically between half a bit and two bits less skillful than a LAP system based on the same model; can a multi-model forecast system regain some of this potential skill? Figure~\ref{fig:Weights_small} shows that this is unlikely, as the determination of model-weights given SAP varies tremendously relative to their LAP values.
Again, it is the performance of the combination of weights that determine the skill of the forecasts, so this variation need not always deadly.

\begin{figure}[!h]
\hbox{
  \epsfig{file=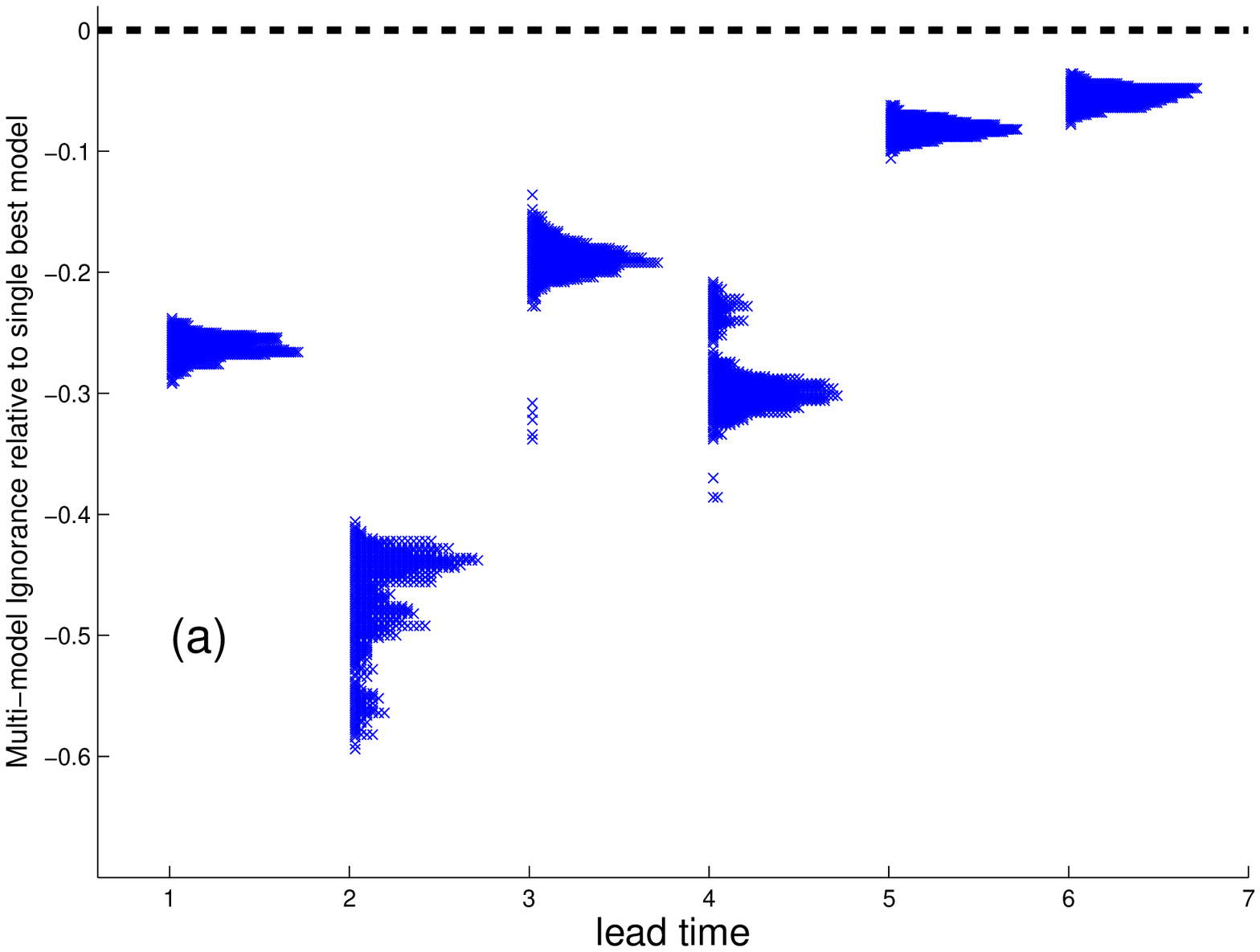, width=0.32\columnwidth, height=5cm}
  \epsfig{file=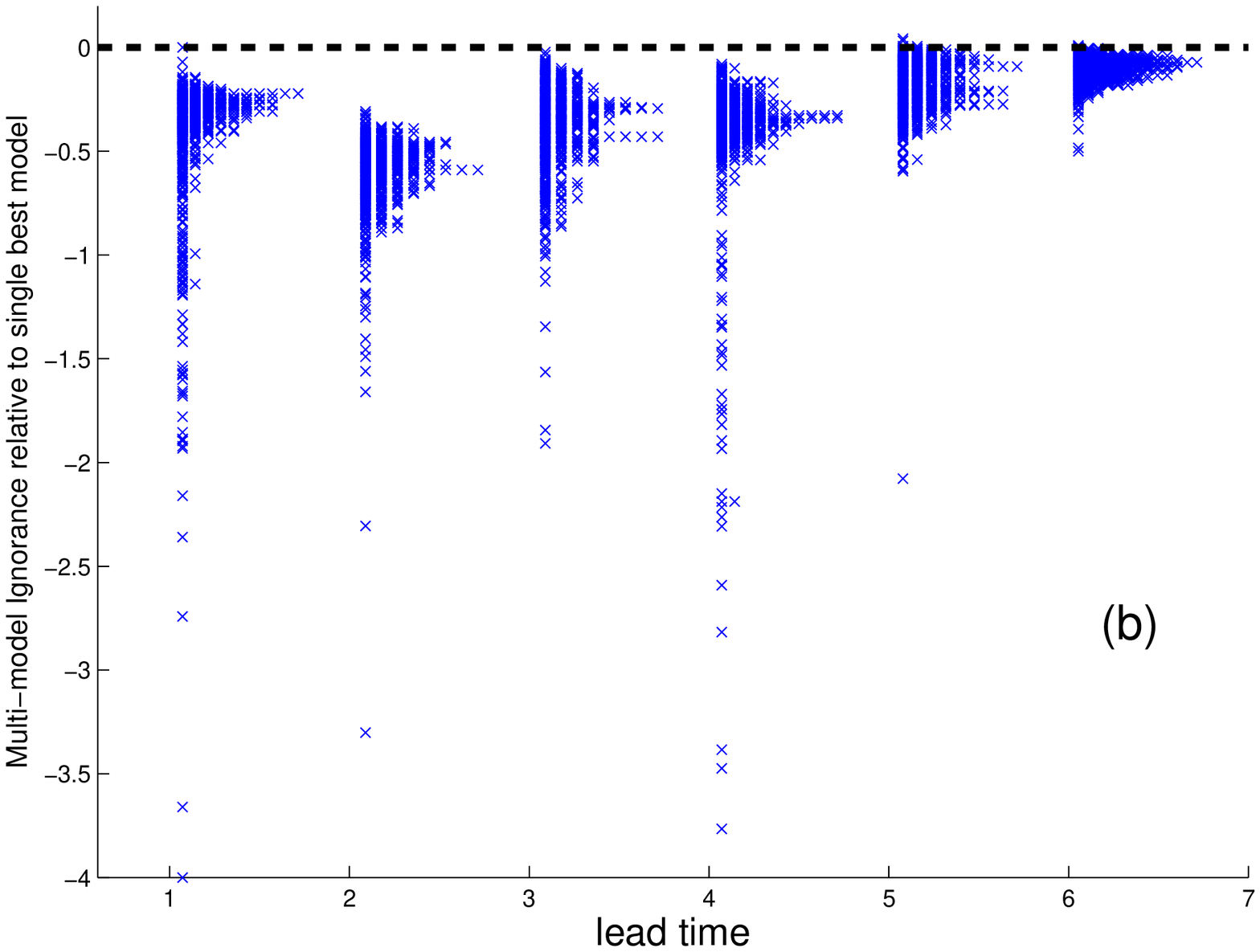, width=0.32\columnwidth, height=5cm}
  \epsfig{file=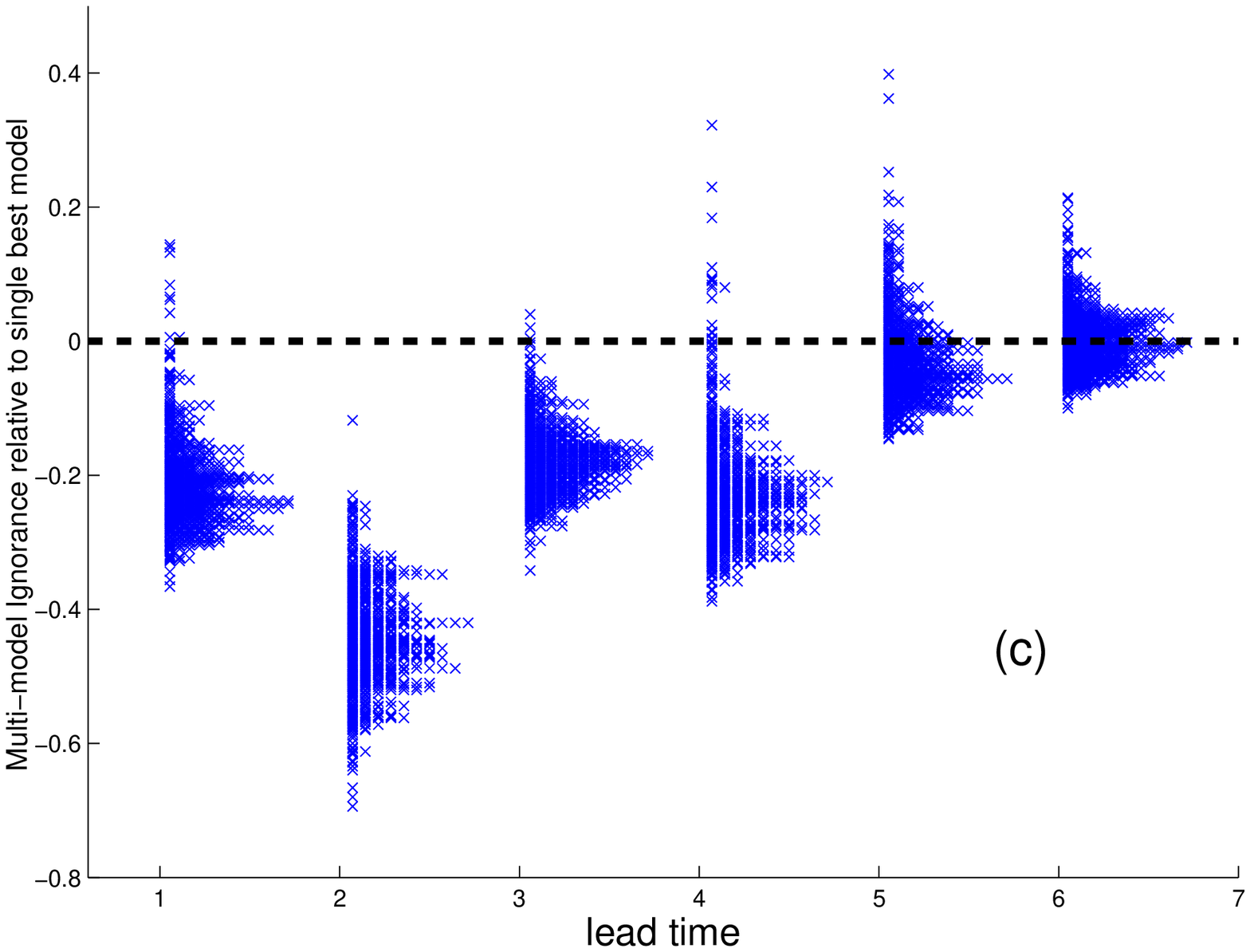, width=0.32\columnwidth, height=5cm}
  
}
\caption{Ignorance of multi-model ensemble relative to the single best model. The blue crosses represent the histogram of Ignorance of multi-model ensemble relative to single best model (black dashed line). (a)  Model weights and dressing and climatology-blend parameter are fitted based on 512 large archives, each contains 2048 forecast-outcome pairs. (b) Model weights and dressing and climatology-blend parameters are fitted based on 512 small archives, each contains 40 forecast-outcome pairs. (c) The Ignorance of the multi-model ensemble is calculated using model weights and dressing and climatology-blend parameter which are fitted based on 512 small archives, while the Ignorance of the single best model is calculated based on 512 large archives.}
\label{fig:MME_BM}
\end{figure}

Figure~\ref{fig:MME_BM} shows skill of the multi-model systems relative to a system based on single best model. Both SAP and LAP forecast systems show the multi-model system usually outperforms the single model. Comparing SAP multi-model systems with the single best model SAP system (Figure~\ref{fig:MME_BM}b), the advantage of the multi-model system(s) is stronger when the best model (as well as all the parameters: model weights and dressing and climatologicy-blended parameters) are ill-identified. Comparing SAP multi-model systems with the single best model LAP system (Figure~\ref{fig:MME_BM}c), however, the advantage of the multi-model system(s) is weaker; multi-model systems do {\bf not} always outperform the single best model, especially at longer lead times. 

At this point, one faces questions of resource distribution, a fair comparison of the multi-model forecast system would be against a single model with n-times larger ensemble. (this, of course, ignores the operational fact that it is much more demanding to maintain an ensemble of models than to maintain a large ensemble under one model.) Secondly, note that for each model $\kappa$ was a function of lead time, at the cost of making ensemble members non-exchangeable one could draw ensembles from distinct groups, and weight these members differently for each lead time. Finally, one could develop methods which treat the raw ensemble members from each of the models as non-exchangeable and use a more complex interpretation to form the forecast.
While the simple forecast framework of this paper is an ideal place to explore such questions, they lie beyond the scope of this paper. Instead, it is concluded by considering the extent to which the multi-model forecast system is more misleading than the single model systems.

\section{Discussion and Conclusions}

A significant challenge to the design of seasonal probability forecasting has been discussed and illustrated in a simple system where multiple models can easily be explored in long time limits. There is no statistical fix to challenges of ``lucky strikes" when a generally poor model places an ensemble member near an outcome by chance, and that particular outcome was not well predicted by the other forecast systems. Similarly ``hard busts" in a small archive can distort the parameters of the forecast systems based on it: when an outcome occurs relatively far from each ensemble member, wider kernels and/or heavier weighting on the climatology results. This may be due to structural model failure, or merely to ``rare" event, where rare would be related to the ensemble size.

In short, the brief duration of the forecast-outcome archive, typically less than 40 years in seasonal forecast, limit the clarity both with which probability distributions can be derived from individual models and with which model weights can be determined. No clear solution to this challenge has been proposed, and while improvements on current practice can be made, it is not clear that this challenge can be met. Over long periods, like the 512 years, the climate may not be well approximated as stationary; in any event both observational systems and the models themselves will evolve significantly, perhaps beyond recognition.

One avenue open to progress is in determining the relative skill of ``the best model" (or a small subset) and the full diversity of models. Following~\cite{Brocker08} it is argued that a large ensemble, forecast system under the best model, may well outperform the multi-model ensemble forecast system when both systems are given the same total computer power; to test this in practice requires access to larger ensembles under the best model.

A second avenue is to reduce the statistical uncertainty of model fidelity within available archive. This can be done by running large ensembles (much greater than ``9", indeed greater than might be operationally feasible) under each model. This would allow identification of which models have significantly different probabilities distributions, and the extent to which they are (sometimes) complementary. Tests with large ensembles also reveal the ``bad busts" due to small ensemble size to be what they are; it can also suggest that those which remain are indeed due to structural model error.

It is suggested that perhaps the most promising way forward is to step away from the statistics of the ensembles, and example the physical realism of the individual trajectories. One can look for shadowing trajectories in each model, and one can attempt to see what phenomena limit the models ability to shadow. Identifying these phenomena, and those that cause them in turn, would allow model improvement independent of the probabilistic skill of ensemble systems. This approach is not new of course, but the traditional physical approach to model improvement which dates back to Charney. Modern forecasting methods do offer some new tools~\cite{Judd08}, and the focus on probabilistic forecasting is well placed in terms of prediction; the point here is merely that probabilistic forecast skill, while a sharp tool for decision support, may prove a blunt tool for model improvement when the data are precious.

\begin{center}
      {\bf APPENDIX}
    \end{center}

\begin{appendix}
\appendix
\section{From Simulation to a Predictive Distribution}

An ensemble of simulations is transformed into a probabilistic distribution function by a combination of kernel dressing and blending with climatology (see \cite{Brocker08}). An $N$-member ensemble at time $t$ is given as $X_{t}=[x^{1}_{t},...,x^{N}_{t}]$, where $x^{i}_{t}$ is the value of a observable quantity for the $i^{th}$ ensemble member. For simplicity, ensemble members under given a model are considered exchangeable. Kernel dressing defines the model-based component of the density as:
\begin{eqnarray}
 \label{eq:SKD}
 p(y:X,\sigma)=\frac{1}{N\sigma}\sum^{N}_{i}K\left(\frac{y-(x^{i})}{\sigma}\right),
\end{eqnarray}
\noindent where $y$ is a random variable corresponding to the density function $p$ and $K$ is the kernel, taken here to be
\begin{eqnarray}
 \label{eq:SKD1}
  K(\zeta)=\frac{1}{\sqrt{2\pi}}exp(-\frac{1}{2}\zeta^{2}).
\end{eqnarray}
\noindent Thus each ensemble member contributes a Gaussian kernel centred at $x^{i}$. For a Gaussian kernel, the kernel width $\sigma$ is simply the standard deviation determined empirically as discussed below.

{For any finite ensemble, there remains the chance of $\sim\frac{2}{N}$ that the outcome lies outside the range of the ensemble even when the outcome is selected from the same distribution as the ensemble itself. Given the nonlinearity of the model, such outcomes can be very far outside the range of the ensemble members. In addition to $N$ being finite, the simulations are not drawn from the same distribution as the outcome as the ensemble simulation system is not perfect. To improve the skill of the probabilistic forecasts, the kernel dressed ensemble may be blended with an estimate of the climatological distribution of the system (see~\cite{Brocker08} for more details, {\cite{Roulston03} for an alternative kernels} and~\cite{Raftery05} for a Bayesian approach).} The blended forecast distribution is then written as
\begin{eqnarray}
 \label{eq:blending}
    p(\cdot)=\alpha p_{m}(\cdot)+(1-\alpha)p_{c}(\cdot),
\end{eqnarray}
\noindent where $p_{m}$ is the density function generated by dressing the model ensemble and $p_{c}$ is the estimate of climatological density. {The blending parameter $\alpha$ determines how much weight is placed in the model. Specifying both values (kernel width $\sigma$, and climatology blended parameter $\alpha$) at each lead time defines the forecast distribution. }Both parameters are fitted simultaneously by optimising the empirical Ignorance score in the training set.


\end{appendix}

\section*{Acknowledgment}
This research was supported by the LSE's Grantham Research Institute on Climate Change and the Environment and the ESRC Centre for Climate Change Economics and Policy, funded by the Economic and Social Research Council and Munich Re; it was also funded as part of the EPSRC-funded Blue Green Cities (EP/K013661/1). Additional support for H.D. was also provided by the National Science Foundation Award No. 0951576 ``DMUU: Center for Robust Decision Making on Climate and Energy Policy (RDCEP)". L.A.S. gratefully acknowledges the continuing support of Pembroke College, Oxford.

\end{document}